\documentclass[12pt]{article}
\usepackage{amsmath}
\usepackage{amsfonts}
\usepackage{amssymb}
\usepackage{cite}

\def\hybrid{\topmargin -20pt    \oddsidemargin 0pt
        \headheight 0pt \headsep 0pt
        \textwidth 6.25in       
        \textheight 9.5in       
        \marginparwidth .875in
        \parskip 5pt plus 1pt   \jot = 1.5ex}

\hybrid

\newcommand{\nn}{\nonumber}
\newcommand{\ax}{\alpha}
\newcommand{\bx}{\beta}

\newcommand{\dx}{\delta}
\newcommand{\ox}{\omega}
\newcommand{\Ox}{\Omega}

\newcommand{\Lx}{\Lambda_1}
\newcommand{\awedge}{\wedge^\ast \!}
\newcommand{\awg}{\wedge^\ast \!}
\newcommand{\wg}{\wedge}
\newcommand{\G}{\mathcal{G}}
\newcommand{\del}{\partial}
\newcommand{\N}{\mathcal{N}}
\newcommand{\IM}{\textrm{Im} \,}
\newcommand{\RE}{\textrm{Re} \,}
\newcommand{\CY}{\textrm{Calabi-Yau }}

\newcommand{\K}{\mathcal{K}}
\newcommand{\F}{\mathcal{F}}
\renewcommand{\L}{\mathcal L}
\newcommand{\Mc}{\mathcal M} 
\newcommand{\cM}{\mathcal M} 

\def\theequation{\arabic{section}.\arabic{equation}}

\begin{document}

\begin{titlepage}
\begin{center}

\hfill hep-th/0202168\\

\vskip 3cm
{\large \bf  Type II Theories Compactified on Calabi-Yau
Threefolds in the Presence of Background Fluxes}\footnote{Work
supported by:
DFG -- The German Science Foundation,
GIF -- the German--Israeli
Foundation for Scientific Research,
European RTN Program HPRN-CT-2000-00148 and the
DAAD -- the German Academic Exchange Service.}

\vskip .5in

{\bf Jan Louis and Andrei Micu\footnote{email: {\tt
      j.louis@physik.uni-halle.de, micu@physik.uni-halle.de}}}  \\

\vskip 0.8cm
{\em Fachbereich Physik, Martin-Luther-Universit\"at Halle-Wittenberg,\\
Friedemann-Bach-Platz 6, D-06099 Halle, Germany}

\end{center}

\vskip 2.5cm

\begin{center} {\bf ABSTRACT } \end{center}

\noindent
Compactifications of type II theories on \CY threefolds including
electric and magnetic background fluxes are discussed. 
We derive the bosonic part of the four-dimensional low energy effective action 
and show that it is a non-canonical $N=2$ supergravity 
which includes a massive two-form. The symplectic invariance of the theory
is maintained as long as the flux parameters transform as
a symplectic vector and a massive two-form which couples to both electric and
magnetic field strengths is present.
The mirror symmetry between 
type IIA and type IIB compactified on mirror manifolds is 
shown to hold for R-R fluxes at the level of the effective action.
We also compactify type IIA in the presence of
NS three-form flux but the mirror symmetry in this case  remains unclear.

\vfill
February 2002 

\end{titlepage}

\section{Introduction}
\setcounter{equation}{0}

Calabi-Yau compactifications of heterotic and type II  theories
have been studied intensively in the past 
since they lead to consistent string theories below the critical
dimension $d=10$. In particular compactifications
on Calabi-Yau threefolds $Y_3$ result in four flat Minkowskian
space-time dimensions $(d=4)$ and a low number of 
unbroken supersymmetries. 
Their effective theories are 
supergravities coupled to a set of vector- and matter multiplets
with $N=1$ supersymmetry in the case of the heterotic string
and $N=2$ supersymmetry for type II strings. 
The low energy effective theories 
all share the feature that they contain
a (large) number of gauge neutral moduli multiplets which are
flat directions of the effective potential and thus parameterize
the vacuum degeneracy of the theory.

Generalization of Calabi-Yau compactifications are
possible if one allows a $p$-form field strength $F_p$ to take 
a non-trivial background value $e_I$ along appropriate cycles $\gamma_I$
in the compact Calabi-Yau 
manifold, i.e.\ $\int_{\gamma_I} F_p = e_I$.  
Depending on the choice of these background fluxes the 
metric is deformed
and the direct product of a  four-dimensional Minkowskian
space times a Calabi-Yau threefold is replaced by a warped product 
\cite{AS1,dWS}. This generically introduces
a potential for the moduli and turns an ordinary supergravity
into a gauged or massive supergravity.
The consistency of such generalized compactifications was discussed in
\cite{SS,BRGPT,CLPST,LLP} while various other aspects 
have been studied previously in
 refs.\ \cite{AS1,dWS,RW,Bac,PS,AGNT,JM,TV,PM1,CKLT,GKP,CKKL,GD,LM,CKL}.

The background fluxes $e_I$ are quantized in units 
of the string scale \cite{RW,PS} and thus do not represent a 
continuous deformations of the Calabi-Yau compactification.
However, in the low
energy supergravity they do appear as continuous parameters and hence can
be
discussed as continuous deformations of the well known low energy
effective
theories derived for vanishing flux background \cite{PS}. 
In the gauged
supergravity the flux parameters play the role of masses and gauge
charges.

The purpose of this paper is to perform a Kaluza-Klein
reduction of the ten-di\-mensional type II supergravities
on compact Calabi-Yau threefolds with all possible
background fluxes turned on.
We derive the bosonic part of the resulting low energy
effective action and show - whenever possible - its consistency with
gauged supergravity.\footnote{For the heterotic
string
we performed a similar analysis in ref.\ \cite{LM}.} 
We find that if magnetic charges are turned on 
a two-form becomes massive and the resulting $N=2$ supergravity
is not easily related to the known $N=2$ supergravities \cite{wp,bw,N=2}.
We show that the symplectic invariance of the ungauged
$N=2$ supergravity continues to hold 
as long as the background flux parameters transform as a symplectic vector.
This leads to a  symplectically invariant potential which 
for compactifications of type IIB theories was previously derived 
in refs.\ \cite{JM,TV,PM1}. However, the 
issue of the symplectic invariance was not resolved and 
here we show that the presence of a massive
two-form is crucial for the symplectic invariance of the theory.
We believe that this is a more general feature of gauged supergravity
and that the existing  $N=2$ supergravities have to be amended
by including the possibilty of massive two-forms.
Only in this more general framework a symplectically invariant theory can
arise. 

A second aspect of our paper concerns the mirror
symmetry between type IIA compactified on $Y_3$ and type IIB
compactified on the mirror manifold $\tilde Y_3$.
This duality holds for vanishing background fluxes but
its validity in the presence of fluxes
is unclear. We show that for the background fluxes of 
Ramond-Ramond (R-R) $p$-forms 
mirror symmetry holds at the level of the effective theories
while for background fluxes in the Neveu-Schwarz (NS) sector
it is not easily established \cite{TV,CV,CKKL,CKL}. A similar problem
arises for the non-perturbative dualties 
relating type IIA compactified on $K3$ to heterotic on $T^4$ \cite{HLS}
and type IIA compactified on $Y_3$ to heterotic on $K3\times T^2$ 
\cite{CKKL,LM}.\footnote{An interesting suggestion for its cure has
been put forward in refs.\ \cite{BBRS,HS2}.}

This paper is organized as follows. In section \ref{RRf} we focus on type 
IIA supergravity compactified on Calabi-Yau threefolds $Y_3$ in the
presence of background fluxes. It turns out that in order to establish 
the mirror symmetry to type IIB compactifications it is necessary
to start from the $N=2$ massive version of the ten-dimensional type IIA
supergravity where the NS two-form $B_2$ is massive \cite{Rom}. 
We first briefly recall this theory including its symmetry properties.
In section 2.1 we then perform the compactification
on $Y_3$ turning on the $2 h_{1,1}$ possible background values of
the R-R two-form field strength $\hat F_2$ and the R-R four-form
field strength $\hat F_4$. ($h_{1,1}$ is the Hodge number
of the cohomology group $H^{1,1}(Y_3)$.)
Furthermore, the four-dimensional low energy effective theory includes 
a three-form (with a four-form field strength) which in $d=4$
is Poincar\'e dual to a constant. This constant is 
an additional parameter of the theory so that together with the 
ten-dimensional mass parameter of $B_2$ 
the theory depends on $2 h_{1,1}+2$ `fluxes'.
The resulting low energy effective action is not
a standard $N=2$ gauged supergravity in that also in $d=4$
the two-form $B_2$ is generically massive. The mass terms depend
on the magnetic flux parameters and vanish for purely
electric fluxes.
As far as we know this situation 
has not been discussed previously
in the supergravity literature.

In section 2.2 we discuss the gauge invariance of this massive theory
and furthermore show that 
as long as the flux parameters are appropriately transformed,
the equations of motions are invariant
under a generalized electric-magnetic duality 
which is part of an $Sp(2 h_{1,1}+2)$ transformation.
To establish this symmetry it is crucial to start from the
massive type IIA theory in $d=10$ and to also include 
the constant dual of the three-form.
A similar observation was made in \cite{HLS} 
where $K3$ compactifications of type IIA in $d=6$ were studied.
The perturbative $SO(4,20)$ T-duality which is present
in the absence of background fluxes can only be maintained
if one starts from the massive ten-dimensional IIA
theory and furthermore  transforms the flux parameter  in the vector
representation of $SO(4,20)$. 
Nevertheless the appearance of the symplectic invariance 
in the $d=4$ theory is somewhat unexpected since it
is usually lost in gauged supergravities.

In section 2.3 we discuss the relation with the standard gauged 
$N=2$ supergravities. By fixing the symplectic
invariance one can go to a particular gauge where all magnetic
charges vanish  and $B_2$ remains massless.
We show that in this gauge the theory is a special case
of the known $N=2$ gauged supergravities.
However, in an arbitrary gauge $B_2$ is massive and
the corresponding supergravity couplings are not independently known.
Nevertheless, in $d=4$ a massive two-form is dual to a massive 
one-form \cite{Tap,QuT,SmS}
and therefore one might suspect that in the dual formulation
consistency with the standard gauged supergravity is achieved.
In section~\ref{relgs} (and appendix~\ref{Mdual}) we explicitly 
perform the duality transformation and show that also
in the dual basis the gauged supergravity is non-canonical.

In section 2.4 we turn our attention to non-trivial fluxes 
of the NS three-form $\hat H_3$ which have not been discussed
previously in the literature. We derive the low energy effective theory
and establish the consistency with $N=2$ gauged supergravity.
In this case the potential depends non-trivially
on the scalar in the hypermultiplets
whereas the dependence on the scalars in the vector multiplets 
is only via an overall volume factor.
This is exactly opposite to the case of R-R fluxes which induce a potential
for the vector-scalars leaving the hyper-scalars 
(except the dilaton) undetermined.
Furthermore, only the graviphoton participates in the gauging
but no charged states with respect to the other gauge fields appear.

In section 3 we briefly discuss type IIB compactifications
on $Y_3$ with non-trivial R-R three-form flux $\hat F_3$.
This case has been considered previously in refs.\ \cite{JM,TV,PM1,CKLT,GD}
and therefore we keep our presentation short. However, we do
establish the fact that
also in this case $B_2$ becomes massive for non-vanishing magnetic
charges.
With this result  we are able to establish the mirror symmetry
to type IIA compactified on the mirror threefold in the presence
of R-R background fluxes at the level of the low energy
effective theories.

Section 4 contains our conclusions and some of the more
technical aspects of this paper are relegated
to five appendicies.
Appendix A summarizes our notation. Appendix B briefly recalls
$N=2$ supergravity in $d=4$.
Appendix C assembles the necessary facts about
the moduli space of Calabi-Yau threefolds. 
Appendix D redoes the compactification
of massless type IIA on $Y_3$ with special emphasis on the 
role of the R-R three-form $C_3$ and its Poincar\'e dual constant.
Already in this simpler case the constant is an additional
 parameter of the effective theory and turns the ordinary type IIA
supergravity into a gauged supergravity.
Finally appendix E discusses the Poincar\'e dualities
of massless and massive two-forms  and of  three-forms in $d=4$.

\section{Compactification of massive type IIA supergravity on Calabi-Yau
  threefolds with background fluxes} 
\label{RRf}\setcounter{equation}{0}

Let us start by compactifying the ten-dimensional 
massive type IIA theory  \cite{Rom} on a \CY threefold in the presence
of background fluxes. The compactification of ordinary
massless type IIA supergravity on \CY threefolds was performed in ref.\
\cite{BCF} which we briefly recall in appendix~\ref{IIA/CY3}.
The massless modes of the ten-dimensional IIA theory
comprise in the NS-NS sector the metric, 
a two form  $\hat B_2$ and the dilaton 
$\hat \phi$ while the RR-sector contains a
vector field $\hat A_1$ and a three-form $\hat C_3$.
In the massive version $\hat B_2$ is massive and a cosmological constant
is present. The action reads
\cite{Rom}\footnote{This differs from \cite{Rom} by a
  redefinition of the mass parameter $m \to  -m$.}
\begin{eqnarray}
  \label{mIIA}
  S & = & \int \bigg[ e^{-2\hat\phi} \left( \frac12 \hat R ^{\;\ast\!} {\bf
  1} + 2 d \hat\phi \awedge d \hat\phi - \frac14 \hat H_3 \awedge \hat H_3
  \right) \nn \\
  & & \qquad - \frac12  \, \left( \hat F_2 \awedge  \hat  F_2  +  \hat F_4
  \awedge \hat F_4 \right)  +  \L_{top}  - \frac12 m^{2\; \ast \!} {\bf 1}
  \bigg ] \ , 
\end{eqnarray}
where the field strengths are defined as
\begin{equation}
  \label{mfs}
   \hat F_2 = d\hat A_1 + m \hat B_2 \; , \quad  \hat F_4 = d \hat C_3 - \hat B_2 \wg d\hat A_1 -
  \frac{m}{2} (\hat B_2)^2 \; , \quad \hat H_3 = d \hat B_2 \, ,
\end{equation}
and the topological 
terms read 
\begin{eqnarray}
  \label{CSdef}
  \L_{top} & = & - \frac12\Big[\hat B_2 \wg d\hat C_3 \wg d\hat C_3 - (\hat
  B_2)^2 \wg d\hat C_3 \wg  d\hat A_1 + \frac13 (\hat B_2)^3 \wg d\hat A_1 \wg
  d\hat A_1 \nn \\
  & & \qquad \quad - \frac{m}{3} (\hat B_2)^3 \wg d \hat C_3 +
  \frac{m}{4} (\hat B_2)^4 \wg d \hat A_1 + \frac{m^2}{20} (\hat B_2)^5 \Big]
  \, .
\end{eqnarray}
Throughout the paper we use differential form notation 
(summarized in appendix~A) and denote differential
forms in $d=10$ by a hat $\hat {\ }\ $. Furthermore we abbreviate 
$\hat B_2 \wg \hat B_2 = (\hat B_2)^2$ etc.\ .
The  massive IIA supergravity has an unbroken ten-dimensional
$N=2$ supersymmetry and 
in the limit $m \to 0$ the action for ordinary type IIA
theory (recorded in (\ref{S10'})) is recovered.
The action (\ref{mIIA}) is invariant under the following 
three Abelian gauge transformations (with parameters 
$\hat\Theta,\hat\Sigma_2,\hat\Lambda_1$)
\begin{eqnarray}
  \label{ginv}
\delta \hat A_1 &=&  d \hat \Theta \ ,\qquad 
\delta \hat C_3 = d \hat \Sigma_2\ , \nn \\
  \dx \hat B_2 &=& d \hat \Lambda_1 \; , \quad \dx \hat C_3 = \hat \Lambda_1
  \wg d\hat A_1 \ , \quad \dx \hat A_1 = - m \hat \Lambda_1 
\, .
\end{eqnarray}
As already mentioned, in the limit $m \to 0$ we recover the standard type IIA
supergravity and hence the degrees of freedom described by the two
theories is the same. However, in the massive theory
 these degrees of freedom are redistributed in
that $\hat B_2$ describes a massive two-form 
(which carries $
\binom{9}{2}=36$ degrees of freedom) 
while $\hat A_1$ carries no degree of freedom 
since it can be
gauged away using (\ref{ginv}). This is the analog of the unitary gauge
in the standard Higgs mechanism 
where $\hat A_1$ plays the role of the Goldstone boson which  is
eaten by $\hat B_2$. 

Compactification of massive IIA supergravity on a Calabi-Yau threefold $Y_3$
results in an effective theory in $d=4$ with $N=2$ supersymmetry and  
proceeds as in the
massless type IIA case (cf. appendix \ref{IIA/CY3}). 
The ten-dimensional metric gives rise
to $h_{1,1} + 2h_{1,2}$ scalar fields related to 
$h_{1,1}$ K\"ahler deformations $v^i, i=1,\ldots,h_{1,1}$
and the $h_{1,2}$ (complex) deformations $z^a, a= 1,\ldots,h_{1,2}$
of the complex structure.\footnote{A summary of the Calabi-Yau
geometry is assembled in appendix~\ref{CYM}.}
The two-form $\hat B_2$ decomposes 
into a two form $B_2$ in $d=4$
and $h_{1,1}$ scalar fields $b^i$ according to
\begin{equation}
\hat B_2 = B_2 + b^i \omega_i\ , \quad i= 1,\ldots,  h_{1,1} \, ,
\end{equation}
 where $\omega_i$ are harmonic $(1,1)$-forms
which form a basis of $H^{1,1}(Y_3)$. The $b^i$ combine with the $v^i$
to form complex fields $t^i = b^i+iv^i$.
The three-form $\hat C_3$ decomposes into a four-dimensional three form $C_3$,
$h_{1,1}$ vector fields $A^i$ and $2h_{1,2}+2$ (real) scalar fields 
$\xi^A, \tilde \xi_A$
\begin{equation}
 \hat C_3 =  C_3 + A^i \wedge \ox_i + \xi^A \ax_A + \tilde\xi_A \bx^A , 
\quad A=0, \ldots, h_{1,2}    \ ,
\end{equation}
where $(\ax_A,\bx^A)$ are 
harmonic three-forms
which form  a real basis of $H^3(Y_3)$ (cf.\  appendix \ref{SGH3}).
Finally the 1-form in $d=10$ only results
in a four-dimensional 1-form i.e.\ $\hat A_1 = A^0$.
Together these (bosonic) fields assemble into the $N=2$ gravity multiplet 
$(g_{\mu \nu}, A^0)$, $h_{1,1}$ vector
multiplets $(A^i,t^i)$, $h_{1,2}$ hypermultiplets
$(z^a, \xi^a, \tilde \xi_a)$ and one tensor multiplet 
$(B_2,\phi, \xi^0, \tilde \xi_0)$.
In $d=4$ a two-form is dual to a scalar and hence the tensor multiplet
can be dualized to an additional (universal) hypermultiplet.
A four-dimensional three form $C_3$ is dual to a constant
and thus the compactified massive type IIA theory 
already contains two parameters, the mass $m$ of the ten-dimensional
action and the constant $e_0$ which is dual to $C_3$.
Both of these constants are related to the cosmological constant
and as we will see shortly they turn the ordinary $N=2$ supergravity
into a gauged supergravity.\footnote{In ref.\ \cite{BCF} the case
$e_0 = 0$ was considered. In appendix \ref{IIA/CY3} we derive
the action of massless type IIA compactified
on $Y_3$ for arbitary $e_0$. The quantization condition on $e_0$ has been 
discussed in refs.\ \cite{BP}.}

\subsection{Turning on R-R fluxes}

Let us now turn to the compactification of massive
type IIA supergravity in the presence of background fluxes
for the RR field strengths $\hat F_2$ and $\hat F_4$. 
This will add $2h_{1,1}$ parameters $(e_i,m^i)$
into the action. We assume
that the fluxes are turned on perturbatively so that the light $d=4$
spectrum  is not modified. More specifically we start from 
the standard reduction Ansatz (cf. appendix \ref{IIA/CY3})
\begin{eqnarray}
  \label{mCYred}
  \hat A_1 & = & A^0 \ ,\qquad
  \hat B_2 =  B_2 + b^i \ox_i\ , \nn \\
  \hat C_3 & = & C_3 + A^i \wedge \ox_i + \xi^A \ax_A + \tilde\xi_A \bx^A\ , 
\end{eqnarray}
and modify 
 the R-R-field strength according to\footnote{%
The minus sign in the last relation was chosen to make the
  symplectic invariance explicit later.}
\begin{equation}
  \label{flcy}
  d \hat C_3 \to d  \hat C_3 +    e_i \tilde \ox^i \ , \qquad
  d  \hat A_1 \to   d  \hat A_1 - m^i \ox_i \ .
\end{equation}
The flux parameters $(e_i,m^i)$  are constants and 
$\tilde \ox^i$ are harmonic $(2,2)$-forms which form a basis
of $H^{2,2}(Y_3)$ and which are dual to the $(1,1)$-forms $\ox_i$,
i.e.\ they obey the normalization of eq.\  (\ref{normH2}). 
Notice that 
we can consistently do this modification at the level of the action since in
(\ref{mIIA}) the fields $\hat A_1$ and $\hat C_3$ appear only through their
Abelian field strengths $d \hat A_1$ and $d \hat C_3$.
In terms of the field strengths defined in (\ref{mfs})
turning on fluxes according to (\ref{flcy}) amounts to  
\begin{eqnarray}
  \label{f4}
  \hat{H_3} & = & d B_2 + d b^i \ox_i\ , \nn \\
  \hat{F}_2 & = & d A^0 + m B_2 - (m^i - m b^i) \ox_i \ ,\nn \\
  \hat{F}_4 & = & d C_3 - B_2 \wg d A^0 - \frac{m}{2} (B_2)^2 +
  (d A^i - d A^0 b^i + m^i B_2 - m B_2 b^i) \wg \ox_i \nn \\
  & & + (d\xi^A \ax_A + d\tilde \xi_A \bx^A)  + (b^i m^j - \frac12 m b^i b^j)
  \K_{ijk} \tilde \ox^k + e_i \tilde \ox^i \, ,
\end{eqnarray}
where $\K_{ijk} =  \int_{Y_3} \ox_i \wg \ox_j \wg \ox_k$ and the last equation
used (\ref{oxstar}).

To derive the four-dimensional effective action we insert (\ref{f4}) into
(\ref{mIIA}). Before giving the final result let us discuss the `new' terms
which arise due to the presence 
of the parameters $(e_0,e_i,m,m^i)$ and are absent in the standard IIA theory.
The kinetic term of $\hat A_1$ gives a contribution to the potential 
  \begin{equation}
    \label{V1}
    V_1 = 2 \K (m^i - m b^i)(m^j -m b^j) g_{ij} \, ,
  \end{equation}
where  $g_{ij}(v) = \frac{1}{4 \K} \int_{Y_3} \ox_i \awg \ox_j$
 is the metric on the space of K\"ahler deformations and 
$\K$ defined in (\ref{K}) 
denotes the volume of $Y_3$.
In addition  the following interaction and mass terms for $B_2$ arise
\begin{equation}
  \label{intB1}
  \delta \L_{\rm int} =  
  - m \, \K B_2 \awg d A^0 -\frac{m^2 \K}{2}\,  B_2 \awg B_2 \, .
\end{equation}
The kinetic term of $\hat C_3$ also contributes to the potential 
  \begin{equation}
    \label{V3}
    V_3 = \frac{1}{8 \, \K} (e_i + b^k m^l \K_{ikl} - \frac12 m b^k b^l
    \K_{ikl} )(e_j + b^m m^n \K_{jmn} - \frac12 m b^m b^n \K_{jmn}) g^{ij} \, , 
  \end{equation}
where $g^{ij} =  4 \K \int_{Y_3} \tilde \ox^i \awg \tilde \ox^j$
is the inverse metric obeying $g_{ij}g^{jk}=\delta^k_i$.
In addition the following interaction terms arise 
\begin{eqnarray}
  \label{intB}
  \delta \L_{\rm int}  &=& - 4 \K (m^i - mb^i)  B_2 \awg 
  \left(d A^j - d A^0 b^j \right) g_{ij}  \nn\\
  &&   - 2 \K (m^i -m b^i) (m^j - m b^j) g_{ij} B_2 \awg B_2  \\
  && - \frac{\K}{2} \big(dC_3 - B_2 \wg d A^0 - \frac{m}{2} (B_2)^2\big) \awg 
 \big(dC_3 - B_2  \wg d A^0 - \frac{m}{2} (B_2)^2\big) \, . \nn
\end{eqnarray}
Finally the topological terms  (\ref{CSdef}) 
are modified according to
\begin{eqnarray}
  \label{CSm}
  \dx \L_{top} & = & - 
  B_2 \wg \left(dA^i e_i  + b^i dA^j
  m^k \K_{ijk} - b^i e_i d A^0 - b^i b^j m^k \K_{ijk} d A^0\right) \\ 
  & & - \frac12 (2 b^i e_i + b^i b^j m^k \K_{ijk} - \frac{m}{3} b^i b^j b^k
  \K_{ijk}) \, dC_3 + \frac{m}{2} B_2 \wg (d A^i - d A^0 b^i) b^j b^k \K_{ijk}
  \nn \\  
  & &  - \frac12 ( m^i e_i - m b^i e_i + b^i m^j m^k \K_{ijk} - \frac{3m}{2}
  b^i b^j m^k \K_{ijk} + \frac{m^2}{2} b^i b^j b^k \K_{ijk} ) (B_2)^2 \, . \nn 
\end{eqnarray}

To arrange the above expressions in the form of a the standard gauged $N=2$
supergravity we are going to proceed as in appendix \ref{IIA/CY3}. First we
dualize the three-form $C_3$ to a constant. Collecting all couplings of
$C_3$ we obtain 
\begin{eqnarray}
  \label{SA3}
  \L_{C_3}& = & - \frac{\K}{2} \big(dC_3 - B_2 \wg d A^0 - \frac{m}{2}
  (B_2)^2 \big) \awg \big(dC_3 - B_2 \wg d A^0 - \frac{m}{2} (B_2)^2
  \big) \nn \\ 
  & & - 
  \big(b^i e_i + \frac12 b^i b^j m^k \K_{ijk} - \frac{m}{6} b^i b^j b^k
  \K_{ijk} \big)\,  d C_3 \ . 
\end{eqnarray}
The dualization of a three-form in $d=4$ is summarized
in appendix  \ref{3f}. Applying the formulae of
this appendix yields the dual action where the three-form
$C_3$ is traded for a constant $e_0$ 
\begin{eqnarray}
  \label{Se0}
  \L_{C_3}\to  \L_{e_0} & =& - \frac{1}{2 \K} \big(e_0 + e_i b^i + \frac12 b^i b^j m^k
  \K_{ijk} - \frac{m}{6} b^i b^j b^k \K_{ijk}\big)^{2 \; *\!} {\bf 1} \\
  & & - 
  \big(e_0 + e_i b^i + \frac12 b^i b^j m^k \K_{ijk} -
  \frac{m}{6} b^i b^j b^k \K_{ijk}) (B_2\wg dA^0 + \frac{m}{2} (B_2)^2\big)\ . \nn 
\end{eqnarray}
Let us stress that the appearance of the
parameter $e_0$ obtained by dualizing $C_3$
does not depend on the fact that we have turned on other fluxes. 
$\L_{e_0}$ does not vanish in the limit $m=m^i=e_j=0$ and thus is 
also present in
the compactification of massless type IIA supergravity without any fluxes
turned on. In appendix \ref{IIA/CY3} we show that $e_0$ becomes the charge
of the scalar $a$ which is dual to $B_2$ and a potential consistent
with the standard $N=2$ gauged supergravity is induced.

Defining the four-dimensional dilaton via $e^{-2\phi} = e^{-2\hat\phi} \, \K$,
using the formulae (\ref{HH}), (\ref{CS}) together with 
(\ref{V1}) -- (\ref{CSm}) and (\ref{Se0})
the low energy effective action takes the form\footnote{%
Strictly speaking also the K\"ahler moduli $t^i$ have to be redefined
by a dilaton dependent factor \cite{BCF}. In order not to overload the
notation we use the same symbol $t^i$ also for the redefined moduli.}
\begin{eqnarray}
  \label{mS4}
  S & = & \int e^{-2\phi} \Big( \frac12 R ^\ast \! {\bf 1} + 2d \phi
  \awedge d \phi - \frac14 H_3 \awedge H_3 - g_{ab} dz^a \awg dz^b 
  - g_{ij} dt^i \awg d{\bar t}^j \Big)   \nn \\
  && + \frac{1}{2}\left(\IM \Mc ^{-1} \right)^{AB} 
  \Big[ d\tilde\xi_A +  \Mc_{AC} d\xi^C \Big] \awg \Big[ d\tilde\xi_B + 
  \bar \Mc_{BD} d\xi^D \Big] \nn\\
  & & +\frac12 H_3 \wg (\tilde\xi_A d \xi^A - \xi^A d \tilde\xi_A ) + \frac12
  \IM \N_{IJ} F^I \awg F^J + \frac12 \RE \N_{IJ} F^I \wg F^J \nn \\ 
  & & - B_2 \wg J_2 - \frac12 M^2\, B_2
  \awg B_2 - \frac12 M^2_T\, B_2 \wg B_2 - V \, ,
\end{eqnarray}
where $I=0,\ldots,h_{1,1}$ and 
$\N_{IJ}, \Mc_{AB}$ are standard supergravity couplings 
defined in (\ref{eq:N}) and
(\ref{N_G}). 
The `new' couplings $J_2, M^2, M^2_T$ 
which depend on the fluxes only appear as couplings
to $B_2$ and are found to be\footnote{%
Note that $M^2$ is positive since in our conventions $\IM \N_{IJ}$
is negative definite.}
\begin{eqnarray}
  \label{MMJ}
  J_2 & = & (e_I F^I -  m^I G_I) \ ,\nn \\
  M^2 & = & - m^I \IM \N_{IJ} m^J \ ,\\
  M^2_T & = & - m^I \RE \N_{IJ} m^J + m^Ie_I\ , \nn 
\end{eqnarray}
where we denoted $m$ by $m^0$ and 
introduced the vectors $m^I = (m^0 , m^i)$, $e_I = (e_0 , e_i)$.
Furthermore, we introduced the magnetic dual of $F^I \equiv d A^I$ by
\begin{equation}
G_I\ \equiv\ \IM {\N_{IJ}} ^{\; *}F^J +  \RE \N_{IJ} F^J\ .
\end{equation}
Finally the string frame potential in (\ref{mS4}) is found to be 
\begin{equation}
  \label{Apot}
  V = -\frac12 (e_I - \bar \N_{IK} m^K) \left( \IM \N \right) ^{-1IJ}
  (e_J - \N_{JL} m^L) \ , 
\end{equation}
where $\left( \IM \N \right) ^{-1}$ is given in (\ref{ImN-1e}).

The action (\ref{mS4}) together with the definitions (\ref{MMJ})
is our first non-trivial result. It gives the low energy effective action 
for massive type IIA supergravity compactified on a Calabi-Yau threefold
in the presence of R-R-background flux. 
As we see the $2h_{1,1}$
flux parameters $(e_i,m^i)$ naturally combine with the mass parameter $m^0$
of the ten-dimensional massive IIA theory and the dual $e_0$ of the
four-dimensional three-form $C_3$ to form the vectors $(e_I,m^I)$.
As we are going to show next these vectors enjoy an
action of a symplectic group $Sp(2h_{1,1}+2)$.
Furthermore the flux parameters introduce a potential $V$,
Green-Schwarz type couplings $B_2 \wedge J_2$, 
a regular and a topological mass term $M,M_T$ for $B_2$.

\subsection{Gauge and symplectic invariance}
\label{syminv}
Let us first focus on the gauge invariance of the action (\ref{mS4}). 
First of all (\ref{mS4}) is manifestly invariant under the standard
one-form gauge transformation  
$\delta A^I = d\Theta^I$. However, the gauge invariance related to the
two-form $B_2$ is less obvious. 
After compactification 
the ten-dimensional gauge transformations (\ref{ginv}) of
the two-form $B_2$ become
\begin{equation}
  \label{gtr4d}
  \dx B_2 = d \Lx \ ,  \quad \dx  C_3 = 
  \Lx \wg d A^0 \ , \quad \dx  A^I = - m^I  \Lx \ .
\end{equation}
As in $d=10$,
the three-form $C_3$ transforms under this gauge transformations. However,
in the dualization of $C_3$ the gauge invariant combination 
$dC_3 - d A^0 \wg B_2 - \frac{m}{2} B_2 \wg B_2$ appeared and 
is dual to the (gauge invariant) constant $e_0$. 
The gauge invariance in the dual action is most easily seen 
by  rewriting  the action (\ref{mS4}) as 
\begin{eqnarray}
  \label{AIB}
  S & = & \int e^{-2\phi} \Big( \frac12 R ^\ast \! {\bf 1} + 2d \phi
  \awedge d \phi - \frac14 H_3 \awedge H_3 - g_{ab} dz^a \awg dz^b 
  - g_{ij} dt^i \awg d{\bar t}^j \Big)   \nn \\
  && + \frac{1}{2}\left(\IM \Mc^{-1} \right)^{AB} 
  \Big[ d\tilde\xi_A + \Mc_{AC} d\xi^C \Big]
  \awg  \Big[ d\tilde\xi_B + \bar \Mc_{BD} d\xi^D \Big]  \, \\
  & & +\frac12 
  H_3 \wg (\tilde\xi_A d \xi^A - \xi_A d \tilde\xi_A )
  + \frac12 \IM \N_{IJ} \check F^I \awg \check F^J 
  + \frac12 \RE \N_{IJ} \check F^I \wg \check F^J \nn \\
  & & -\frac12 B_2 \wg  (\check F^I + dA^I) e_I  - V \ ,\nn
\end{eqnarray}
where we defined
\begin{equation}
\check F^I \equiv dA^I + m^I B_2 \ .
\end{equation}
Under the transformations  
\begin{equation}
  \label{gtrf}
  \dx B_2 = d \Lx \ , \quad \dx  A^I = - m^I  \Lx \, ,
\end{equation}
$\check F^I$ is invariant and it can be easily checked that the action 
(\ref{AIB})
is also not modified.

Let us discuss the symplectic invariance of the theory described by (\ref{AIB}).
The ungauged $N=2$ supergravity is invariant under generalized
electric-magnetic duality transformations which are part
of a symplectic $Sp(2h_{1,1}+2)$ invariance \cite{wp,N=2}.
However, this is not a symmetry of the action, but it leaves the
equations of motion and Bianchi identities invariant.
In gauged supergravity this invariance is generically broken 
as charged states appear and the action is no longer expressed in terms
of only the field strength $F^I$. However, for the case at hand
a symplectic invariance can be maintained as long
as the fluxes $(m^I,e_I)$ are also transformed. 
To see this consider the Bianchi identities and the equations of motions derived
from the action (\ref{AIB})
\begin{eqnarray}
  \label{seoms}
  d \; d A^I & = & d {\check F}^I - m^I d B_2 = 0\ , \nn \\
  \frac{\partial \L}{\partial A^I} &=&  d {\check G}_I - e_I d B_2 = 0 \ , \\
  \frac{\partial \L}{\partial B_2} &=& \frac12 d(e^{-2\phi \; * \!} d B_2)  \! +  \!
  m^I {\check G}_I - e_I {\check F}^I = 0\ , \nn 
\end{eqnarray}
where
\begin{equation}
  \label{FtilA}
  {\check G}_{I} \equiv \RE \N_{IJ} {\check F}^J 
+ \IM \N_{IJ}
\rule{0pt}{10pt}^*{\check F}^J   \, .
\end{equation}
These equations are invariant under the 
symplectic transformations given in (\ref{FGdual}) with 
$(m^I, e_I)$ and $({\check F}^I, {\check G}_I)$ transforming as 
symplectic vectors
\begin{equation}
  \left(\begin{array}{c} m^I\\ e_I \end{array}\right) \to
  \left(\begin{array}{cc} U & Z \\[1mm] W & V \end{array}\right)
  \left(\begin{array}{c} m^I\\ e_I \end{array}\right) \ , \qquad
  \left(\begin{array}{c} \check F^I \\ \check G_I \end{array}\right) \to
  \left(\begin{array}{cc} U & Z \\[1mm] W & V \end{array}\right)
  \left(\begin{array}{c} \check F^I\\ \check G_I \end{array}\right) \ ,
\end{equation}
where $U,V,W,Z$ obey (\ref{spc2}).
Similarly one checks the invariance of $V$ given in (\ref{Apot}).

The symplectic invariance of the equations (\ref{seoms}) is our second 
non-trivial result. In particular it shows that the symplectic invariance of
the potential as observed in \cite{TV,PM1} 
has its deeper origin in the symplectic invariance of (\ref{seoms}).
However, the supergravity which displays this invariance
is not the canonical one but instead features a massive
two-form $B_2$ with very specific couplings to the gauge fields.
It would be interesting to investigate this situation in more detail
from a purely supergravity point of view
without any reference to a flux background of string theory.

\vskip 30pt

\subsection{Relation with gauged supergravity}
\label{relgs}
Let us now investigate the relation between the action derived in
(\ref{mS4}) or (\ref{AIB}) and the standard gauged $N=2$ supergravity
as summarized in appendix~\ref{sg}.
The new ingredients in the action (\ref{mS4}) are the mass terms for $B_2$.
Let us first observe that they all vanish for $m^I=0$. 
Since we have established the symplectic invariance of the theory
we can always do 
a symplectic transformation on the vector 
$(m^I,e_I)$ and go to a basis where all
$m^I$ vanish\footnote{%
We should stress again that from a pure supergravity point of view the fluxes
$m^I, \ e_I$ are just continuous parameters and so there always exist an
$Sp(2h_{1,1} + 2, {\bf R})$ transformation such that the rotated magnetic
fluxes vanish. In a quantum theory however, the fluxes become quantized and
the $Sp(2h_{1,1} + 2, {\bf R})$ invariance is generically broken to
$Sp(2h_{1,1} + 2, {\bf Z})$. In this case, it is impossible to set the
magnetic charges to zero by an $Sp(2h_{1,1} + 2, {\bf Z})$ rotation.}
\begin{equation}
  \label{elch}
  \left(\begin{array}{c} m^I\\ e_I \end{array}\right) \to 
  \left(\begin{array}{c} 0 \\ e'_I \end{array}\right) \ .
\end{equation}

\noindent
In this basis the `new' couplings considerably simplify
and from (\ref{MMJ}) and (\ref{Apot}) one immediately obtains
\begin{equation}
  \label{m0case}
  M=M_T=0\ ,\qquad  J_2 = e'_I F^I \ , \qquad 
  V =  -\frac12 e'_I \left( \IM \N' \right) ^{-1IJ} e'_J \ ,
\end{equation}
where the prime indicates the rotated basis.
The drawback of this basis is that also 
the gauge couplings $\N$  of the
action (\ref{mS4}) change according to  (\ref{nchange})
and the relation to the prepotential as given in (\ref{Ndef})
is more complicated.
So we have the choice to work either with the standard 
gauge couplings and a set of complicated interactions of $B_2$ 
or to transform to a new basis where the gauge couplings are more
complicated but $B_2$ remains massless. 
In this latter basis the consistency with gauged supergravity
is easily established so let us first discuss this case.

For $m^I=0$ $B_2$ is massless and thus can
be dualized to a scalar $a$ as in appendix~\ref{Bdual}.
After a Weyl rescaling $g_{\mu \nu} \to e^{2\phi} g_{\mu \nu}$
the dual action reads  
\begin{eqnarray}
  S &=&  \int \Big[ \frac12 R ^* {\bf 1} 
  + \frac{1}{2}\, \IM\N'_{IJ} F^{\prime I}\awg F^{\prime J}
  + \frac{1}{2} \, \RE\N'_{IJ} F^{\prime I} \wg F^{\prime J}\nn\\
&& \qquad\qquad -  g_{ij} dt^i \awg d
  {\bar t}^j - h_{uv} Dq^u \awg Dq ^v - V_E
  \Big] \ ,
\end{eqnarray}
where 
\begin{eqnarray}
  \label{qkt1}
  h_{uv} Dq^u \awg Dq ^v & = & d\phi \awg d\phi + g_{ab} dz^a \awg dz^b\\
  & +& \frac{e^{4\phi}}{4} \, \Big[ Da + 
  (\tilde\xi_A d \xi^A-\xi^A
  d\tilde\xi_A) \Big] \wg \rule{0pt}{12pt}^* \! \Big[ Da + 
  (\tilde\xi_A d \xi^A-\xi^A d\tilde\xi_A) \Big]\nn  \\ 
  & -& \frac{e^{2\phi}}{2}\left(\IM \Mc^{-1} \right)^{AB} \Big[ d\tilde\xi_A +
  \Mc_{AC} d\xi^C \Big] \awg  \Big[ d\tilde\xi_B + \bar \Mc_{BD} d\xi^D \Big]
  \, , \nn
\end{eqnarray}
and 
\begin{equation}
  \label{elcda}
D a = d a + 2 e'_I A^{\prime I}\ .
\end{equation}
The covariant derivative of $a$ arises from the Green-Schwarz
type interaction $B_2 \wg J_2$ in (\ref{mS4}) and as a consequence
$a$ couples like a Goldstone boson and is charged 
under an Abelian gauge symmetry with gauge charges $e'_I$.
$V_E$ represents the potential in the Einstein frame and is given by
\begin{equation}
  \label{ApotE}
  V_E  = -\frac{e^{4\phi}}{2} e'_I \left( \IM \N' \right)^{-1IJ} e'_J \, .
\end{equation}

In  ref.\ \cite{FeS}
it was shown that $ h_{uv}$ of (\ref{qkt1}) is a quaternionic metric 
in accord with the constraints of $N=2$ supergravity
that the scalars in the hypermultiplets span a quaternionic manifold.
In order to establish the further consistency with gauged $N=2$ supergravity
we need to show that the potential (\ref{ApotE}) is consistent with the
general form of the potential (\ref{pot}) of gauged supergravity.
Let us first note that only one scalar $a$ in the hypermultiplets
carries gauge charge while the scalars $t^i$ in the vector multiplets
remain neutral. In terms of the Killing vectors defined in (\ref{kdef})
and (\ref{gaugeco}) eq.\ (\ref{elcda})  implies 
\begin{equation}\label{killspec}
k_I^u= - 2 e'_I \delta^{ua}\ , \qquad k_I^i =0\ .
\end{equation}
Inserted into (\ref{pot}) using (\ref{qkt1}) one arrives at 
\begin{equation}
  \label{potN2}
  V_E = - \frac12 \left[(\IM \N')^{-1} \right]^{IJ} P^x_I P^x_J 
  + 4 e^K X^I \bar X^J( e^{4\phi} e'_I e'_J - P^x_I P^x_J)\ .
\end{equation}
We are left with the computation of the Killing prepotentials 
$P_I^x$ defined in (\ref{killingpre}). Following ref.\  \cite{JM}
one first observes that for the constant (field independent) 
Killing vectors as in (\ref{killspec}) eqs.\ (\ref{killingpre})
are solved by
\begin{equation}\label{Pomega}
P_I^x = k_I^u \omega_u^x \ ,\quad x=1,2,3\ ,
\end{equation}
where $\omega_u^x$ is the $SU(2)$ connection on the quaternionic manifold.
For the case at hand $\omega_u^x$  has been computed
in \cite{FeS} and here we only need their result 
$\omega_a^x = \frac12 e^{2 \phi} \dx^{3x}$.  Inserted into (\ref{Pomega}) using
(\ref{killspec}) we obtain
\begin{equation}
  \label{kpre}
  P_I^1 = P_I^2 = 0 \, , \quad P_I^3 = 
  e^{2\phi} e'_I\ ,
\end{equation}
which implies $P^x_I P^x_J = h_{uv} k^u_Ik^v_J$. Thus 
the last term in (\ref{potN2}) vanishes 
while the first one
reproduces the potential (\ref{ApotE}).
This establishes the consistency with $N=2$
gauged supergravity. 

Let us return to the discussion of the action in the unrotated basis
where both $e_I$ and $m^I$ are non-zero.
In this case $B_2$ is massive and the relation with the standard gauged
supergravity is not obvious and, as far as we know,
has not been discussed in the literature.
However, one can use the fact that
a massive two-form in $d=4$ is Poincar\'e dual to a massive vector 
\cite{Tap,QuT,SmS}.
This generic duality is briefly summarized in appendix~\ref{Mdual}.
In the following we perform the duality transformation
and display the dual action in terms of only vector fields.

Starting from the action (\ref{mS4}) it is straightforward to apply the
results in appendix~\ref{Mdual}. Denoting by $A^H$ the dual of the
massive $B_2$ the resulting action reads
\begin{eqnarray}
 \label{AIB1}
  S & = & \int e^{-2\phi} \Big( \frac12 R ^\ast \! {\bf 1} + 2d \phi
  \awedge d \phi - g_{ab} dz^a \awg dz^b 
  - g_{ij} dt^i \awg d{\bar t}^j \Big)    \nn \\
  && + \frac{1}{2}\left(\IM \Mc ^{-1} \right)^{AB} 
  \Big[ d\tilde\xi_A +  \Mc_{AC} d\xi^C \Big] \awg \Big[ d\tilde\xi_B + 
  \bar \Mc_{BD} d\xi^D \Big] - V \nn\\
  & &  + \frac12
  \IM \N_{IJ} F^I \awg F^J + \frac12 \RE \N_{IJ} F^I \wg F^J - e^{2\phi} A^H \awg A^H  \nn \\ 
  & & -\frac12 \frac{M^2}{M^4 + M_T^4} \left(F^H - J'_2 \right) \awg
  \left(F^H - J'_2 \right)  \\
  & & + \frac12 \frac{M_T^2}{M^4 + M_T^4} \left(F^H - J'_2 \right) \wg
  \left(F^H -  J'_2 \right)\, ,\nn  
\end{eqnarray}
where
\begin{equation}
  \label{J'2}
 F^H = dA^H\ ,\qquad  J'_2 = J_2 + \frac12 d(\tilde \xi_A d \xi^A- \xi^A
 d\tilde \xi_A ) \, ,
\end{equation}
and the quantities $M , \, M_T$ and $J_2$ are defined in (\ref{MMJ}).
The above action contains an explicit mass term for the vector field $A_H$
 which can equivalently be written as
the covariant derivative of a Goldstone boson
\begin{equation}\label{Aamass}
e^{2\phi} A^H \awg A^H = \frac14 e^{2\phi} D a\awg Da \ ,
\end{equation}
where 
\begin{equation}
  \label{covda}
  D a = d a + 2 A^{\prime H} \, .
\end{equation}
($A^{\prime H}$ denotes the gauge transformed vector potential.)
Inserting (\ref{Aamass}) into (\ref{AIB1}) and absorbing 
$\frac12(\tilde \xi_A d \xi^A- \xi^A d\tilde \xi_A)$ into a further redefinition
of $A^H$ results in 
\begin{equation}
  \label{AIB2}
  S =  \int  \frac12 R ^* {\bf 1} -  g_{ij} dt^i \awg d
  {\bar t}^j - h_{uv} Dq^u \awg Dq ^v
  + \frac{1}{2}\, \IM\hat \N_{\hat I\hat J} F^{\hat I}\awg F^{\hat J}
  + \frac{1}{2} \, \RE\hat \N_{\hat I\hat J} F^{\hat I} \wg F^{\hat J} - V_E \ ,
\end{equation}
where also a Weyl rescaling 
$g_{\mu \nu} \to e^{2\phi} g_{\mu \nu}$ has been performed and we 
introduced the index $\hat I =(I,H)$. $V_E$ is the Weyl rescaled potential
related to $V$ of (\ref{Apot}) by $V_E = e^{4\phi} V$. 
$h_{uv} Dq^u \awg Dq ^v$ is again the standard quaternionic metric 
given in (\ref{qkt1}) with the only difference that (\ref{elcda})
is replaced by (\ref{covda}).
Moreover, the `new' $(h_{1,1} +2)\times (h_{1,1}
+2)$ dimensional gauge coupling matrix $\hat\N_{\hat I\hat J}$ is  given by 
\begin{eqnarray}
  \label{mgcp}
 \hat \N_{IJ} & = & \N_{IJ} - i \, \mu \,  \big(e_I -
  \N_{IK} m^K \big) \big(e_J - \N_{JL} m^L \big) \ , \qquad 
\hat\N_{HH} = - i \,\mu \ ,
\nn \\
\hat\N_{IH} &=&  i \, \mu \, \big(e_I -\N_{IK} m^k
  \big)\ ,   \qquad  \qquad
\mu \equiv \frac{M^2 + i M_T^2}{M^4 + M_T^4}\ .
\end{eqnarray}
One easily shows that $\hat \N_{IJ} m^J = e_I$ and hence 
$\IM\hat \N_{\hat I\hat J}$ has a null vector while $\RE\hat \N_{\hat I\hat
  J}$
has one constant eigenvalue.
This implies that one (linear combination) of the vector fields
only has a topological coupling.\footnote{We thank B.\ de Wit
and S.\ Vandoren for discussions on this point.}

The dualization of $B_2$ resulted in an additional massive 
vector $A^H$ and we chose to write the mass term
as the coupling of a Goldstone boson $a$. The number of physical degrees
of freedom is of course unchanged since the
action (\ref{AIB1})/(\ref{AIB2})  
is still invariant under the gauge transformations
(\ref{gtrf}) which after dualization become
\begin{equation}
  \label{dgtr}
  \dx A^I = - m^I \Lambda_1 \, , \qquad \dx A^H = 0\ .
\end{equation}
$A^H$  being the Poincar\'e dual of
$H_3$ is invariant under (\ref{gtrf})
but one of the other $h_{1,1}+1$ 
vector fields in (\ref{AIB2}) can be gauged away by (\ref{dgtr}).
In this `unitary gauge' the symplectic invariance is lost.
Thus the theory can be formulated in terms of
only vector fields but symplectic invariance demands 
the presence of an additional auxiliary vector field with only
topological couplings.
In any physical gauge the symplectic invariance is broken.

To conclude this section let us discuss another aspect of the
dualization of the massive $B$-field.
Eq.\ (\ref{seoms}) can be solved  for $\check F^I$ and
$\check G_I$ in terms of electric and magnetic potentials $A^I$ and $\tilde
A_I$
\begin{equation}
  \label{empot}
  \check F^I = m^I B_2 + d A^I \ , \quad \check G_I = e_I B_2 + d \tilde A_I
  \, .
\end{equation}
Now the equation of motion for $B_2$ becomes
\begin{equation}
  \frac12 d (e^{-2 \phi} \rule{0pt}{10pt}^* d B_2 ) + m^I d \tilde A_I - e_I d A^I = 0
  \, .
\end{equation}
This suggests that we can introduce a scalar field $a$ (the dual of
$B_2$)  which obeys
\begin{equation}
  \label{emcd}
  e^{-2 \phi} \rule{0pt}{10pt}^* d B_2 = D a \equiv d a - 2 m^I d \tilde A_I +
  2 e_I d A^I \, .
\end{equation}
This definition has the feature that it maintains explicitly the symplectic
invariance closely related to the proposal of \cite{JM,PM1}. 
However, in (\ref{empot}) $B_2$ and not $dB_2$ appears and thus it is not
possible to give an action in terms of the dual scalar $a$ with
electric and magnetic couplings.
Nevertheless, one can compute the electric and magnetic Killing
prepotentials corresponding to the gauging (\ref{emcd}) as suggested in 
\cite{JM,PM1}. They are very similar to the ones found only for the
electrically charged particles (\ref{kpre}) 
\begin{equation}
  \label{emkp}
  P_I^3 = e^{2\phi} e_I , \qquad \tilde P^{I3} 
  = e^{2\phi} m^I , \qquad  P_I^1 =  P_I^2 = \tilde P^{I1} =
\tilde P^{I2} = 0\ .
\end{equation}
Using the formula for the potential suggested in \cite{JM} 
\begin{eqnarray}
  \label{potN2m}
 V_E &=& 4 e^K X^I \bar X^J h_{uv}(k^u_I -\tilde k^{uK} \N_{KI}) 
 (k^v_I -\tilde k^{vK} \bar \N_{KI}) \\
 && - \left[\frac12 (\IM \N)^{-1IJ} + 4e^K X^I \bar X^J   \right]
 (P^x_I - \tilde P^{Kx} \N_{KI}) (P^x_J - \tilde P^{Kx} \bar \N_{KJ} )\ ,\nn 
\end{eqnarray}
which is the
symplectic invariant extension of (\ref{pot}), one immediately recovers the
potential obtained in (\ref{Apot}).

\subsection{NS fluxes}
\label{NSflux}

So far we concentrated on non-trivial flux related to the modification
of the R-R field strength $d\hat C_3$ and $d\hat A_1$.
In this section we discuss the modifications which appear in the type IIA
compactification due to the presence of NS fluxes. For the consistency of
the procedure, we require that the fields which acquire a background value
appear in the action only via the corresponding field strengths. This is easily
achieved for the R-R fields $\hat A_1$ and $\hat C_3$ 
as  can be seen in the form of
the actions (\ref{mIIA}). 
Clearly, for massive type IIA theory,
the NS-NS field $\hat B_2$ cannot appear only via its field strength so we only
have a chance to turn on NS fluxes if we start from the massless type IIA
theory given in (\ref{S10'}). 
However we need to perform the field redefinition
$\hat C_3\to  \hat C_3 + \hat A_1\wedge\hat B_2$
in order to have the action
only depend on the field strength $\hat H_3$ but not $\hat B_2$.
This turns (\ref{S10'}) into 
\begin{eqnarray}
  \label{S10}
  S & = & \int \, e^{-2\hat\phi} \left( \frac12 \hat R ^\ast\! {\bf 1} + 2
  d \hat\phi \awedge d \hat\phi - \frac14  \hat H_3\awedge \hat H_3 \right) \nn \\
  & & - \frac12  \, \left(\hat  F_2 \awedge \hat F_2 + \hat F_4 \awedge \hat
  F_4 \right) + \frac12 \hat H_3 \wedge \hat C_3 \wedge d \hat C_3 \, , 
\end{eqnarray}
where $ \hat F_4 = d \hat C_3 - \hat A_1 \wedge\hat H_3$ and
$ \hat H_3 = d\hat B_2 $. Note that this field redefinition changes the form
of the gauge transformations (\ref{gtrIIA}) which now become
\begin{eqnarray}
  \label{ginv10}
  \dx \hat B_2 & = & d \hat \Lambda_1 \, , 
\qquad \dx \hat C_3\ =\ d \hat \Sigma_2\ , \nn \\
  \dx \hat A_1 & = &  d \hat \Theta \, , 
\qquad  \dx \hat C_3\ =\ \hat \Theta \wg d \hat
  B_2 \, .  
\end{eqnarray}

The compactification of this action proceeds as in appendix~\ref{IIA/CY3}
with the following 
modification of the reduction Ansatz (\ref{CYred})
\begin{equation}
  \label{NSf}
\hat H_3 = H_3 + db^i \ox_i + p^A \ax_A + q_A \bx^A\ ,
\end{equation}
where $H_3=dB_2$ and $(p^A,q_A)$ are the $2h_{1,2}+2$ possible flux parameters 
of $\hat H_3$.
Using (\ref{NSf}) $\hat F_4$ is reduced according to
\begin{equation}\label{NSF}
  \hat F_4 = d C_3 - A^0 \wg H_3 + (d A^i - A^0 d b^i) \wg \ox_i + D \xi^A
  \wg \ax_A + D \tilde\xi_A \wg \bx^A\ ,
\end{equation}
where
\begin{equation}
  \label{NScd}
  D  \xi^A = d  \xi^A - p^A A^0 \ , \qquad 
D \tilde\xi_A = d \tilde\xi_A - q_A A^0 \, .
\end{equation}
Inserting  (\ref{NSf}) and (\ref{NSF}) into the action  (\ref{S10})
results in terms very similar to  (\ref{HH}).
The differences are that the kinetic term  for $\hat C_3$ 
now contains the covariant derivatives $D\xi^A, D\tilde\xi_A$ instead of the
ordinary ones $d\xi^A, d\tilde\xi_A$. 
Furthermore, the kinetic term for $\hat B_2$ (\ref{HH}) now induces a potential
\begin{equation}
\label{VNS1}
V = - \frac14 \, \frac {e^{-2\phi}}{\K}\ (q_A + \Mc_{AC}\, p^C) (\IM \Mc)^{-1AB}
(q_B + \bar \Mc_{BD}\, p^D)\ ,
\end{equation}
where $\Mc$ is defined in (\ref{N_G}).
The last modification due to (\ref{NSf}) appears from the 
topological term in (\ref{S10}) which reads 
\begin{equation}
  \dx \L_{\textrm{top}} = 
  (p^A \tilde\xi_A - q_A \xi^A)\, d C_3\ .
\end{equation}

As before we have to dualize the three-form $C_3$ to a constant. 
Collecting the terms including $C_3$ we find 
\begin{equation}
  \L_{C_3} =   - \frac{\K}{2}\, (d C_3 - A^0 \wedge H_3) \awg 
  (d C_3 - A^0 \wedge H_3) + 
  (p^A \tilde\xi_A - q_A \xi^A)\, d C_3\ .
\end{equation}
By using the formulas of appendix~\ref{3f} we obtain the dual Lagrangian 
\begin{equation}
  \label{VNS2}
  \L_{C_3}\to \L_{e} = -\frac{1}{2 \K} (p^A \tilde\xi_A - q_A \xi^A + e)^2\,
   \rule{0pt}{10pt}^*{\bf 1} + 
   (p^A \tilde\xi_A - q_A \xi^A + e)\, A^0 \wedge H_3 \ ,
\end{equation}
where $e$ is an arbitrary constant, the dual of $C_3$.
The first term in the above expression contributes together with
(\ref{VNS1}) to the scalar potential which reads 
\begin{equation}
  \label{VNS}
  V = - \frac {e^{-2\phi}}{4 \, \K}\ (q + p \Mc) (\IM
  \Mc)^{-1} (q + p\bar \Mc) +\frac{1}{2 \, \K} (p^A \tilde\xi_A - q_A \xi^A +
  e)^2 \ .
\end{equation}

The final thing to do in order to have the compactified action in the standard
form of a gauged supergravity is to dualize $B_2$ to a scalar. 
Collecting the terms including $B_2$ we find 
\begin{eqnarray}
  \L_{B_2} & = & - \frac{e^{-2\phi}}{4} \, H_3 \awg H_3 + \frac12 H_3 \wg
  \Big[\tilde\xi_A d\xi^A -\xi^Ad\tilde\xi_A 
  - 2 (p^A \tilde\xi_A - q_A \xi^A + e)  A^0 \Big]  \\ 
  & = & - \frac{e^{-2\phi}}{4} \, H_3 \awg H_3 + \frac12 H_3 \wg
  \Big[\tilde\xi_A D\xi^A -\xi^AD\tilde\xi_A 
  - (p^A \tilde\xi_A - q_A \xi^A + 2 e) A^0 \Big]\ .\nn 
\end{eqnarray}
Using appendix \ref{Bdual} we obtain the following action  for the dual scalar
$a$ 
\begin{equation}
  \label{duala}
  \L_{B_2}\to \L_a = - \frac{e^{2\phi}}{4} \, \left[ Da + 
    (\tilde\xi_A D\xi^A -\xi^A D\tilde\xi_A ) \right] \wg \rule{0pt}{12pt}^*
  \! \left[ Da + 
  (\tilde\xi_A D\xi^A -\xi^A D\tilde\xi_A) \right] \, , 
\end{equation}
where 
\begin{equation}
  \label{NScda}
  D a = da - 
  (p^A \tilde\xi_A - q_A \xi^A)A^0 - 2 e A^0 \, .
\end{equation}
The final form of the action action is
obtained after going to the Einstein frame and is
similar to the one found in the massless case (\ref{S4f}) 
\begin{equation}
  S =  \int \Big[ \frac12 R ^* {\bf 1} - g_{ij} dt^i \awg d
  {\bar t}^j - h_{uv} Dq^u \awg Dq ^v + \frac{1}{2}\, \IM \N_{IJ} F^I\awg F^{J}
  + \frac{1}{2} \, \RE \N_{IJ} F^I \wg F^J - V_E \Big] ,
\end{equation}
where now 
\begin{eqnarray}
  \label{qktNS}
  h_{uv} Dq^u \awg Dq ^v& = &  d\phi \awg d\phi + g_{ab} dz^a \awg dz^b
  \\  
  & & + \frac{e^{4\phi}}{4} \, \Big[ Da + 
  (\tilde\xi_A D \xi^A-\xi^A D\tilde\xi_A) \Big] \wg \rule{0pt}{12pt}^* \Big[Da + 
  (\tilde\xi_A D \xi^A-\xi^A D \tilde\xi_A) \Big] \nn \\
  & & - \frac{e^{2\phi}}{2}\left(\IM \Mc^{-1} \right)^{AB} 
  \Big[ D\tilde\xi_A + \Mc_{AC} D\xi^C \Big]
   \wg \rule{0pt}{12pt}^* \Big[ D\tilde\xi_B + \bar \Mc_{BD} D\xi^D \Big]  \,
  ,\nn 
\end{eqnarray}
with the covariant derivatives $D \xi^A , \ D \tilde \xi_A$ and $Da$ given in
(\ref{NScd}) and (\ref{NScda}). The Einstein frame potential will only
differ from (\ref{VNS}) by a factor of $e^{4\phi}$ 
\begin{equation}
  \label{NSpotE}
  V_{\textrm{E}} = - \frac {e^{2\phi}}{4 \, \K}\ (q + p \Mc) (\IM
  \Mc)^{-1} (q + p\bar \Mc) +\frac{e^{4\phi}}{2 \, \K} (p^A \tilde\xi_A - q_A \xi^A +
  e)^2 \ .  
\end{equation}
The crucial difference to the previous case of R-R fluxes is 
that now the potential depends on the scalars in the hypermultiplets
but not on the scalars in the vector multiplets.

As before the fluxes turn ordinary supergravity 
into a gauged supergravity 
where a certain isometry of the scalar manifold has been gauged.
The corresponding gauge invariance is
just the compactified version of the 10 dimensional gauge invariance 
(\ref{ginv10}).
Inserted into the compactification Ansatz (\ref{CYred}) modified according
to (\ref{NSf}), the four-dimensional gauge invariance reads
\begin{eqnarray}
  \label{ginv4}
  \dx A^0 & = & d \Theta \ , \qquad \dx d A^i = \Theta \wg d b^i \ ,
  \qquad \dx C_3  = \Theta \wg H_3 \ , \nn \\
  \dx \xi^A & = & p^A  \Theta \ , \qquad \dx \tilde\xi_A = q_A  \Theta\ .
\end{eqnarray}
The covariant derivatives defined in (\ref{NScd}) and 
the action (\ref{duala}) is  invariant under
this transformations provided $a$ transforms according to 
\begin{equation}
  \label{atr}
  a \to a + \big[2 e + 
  (p^A \tilde \xi_A - q_A \xi^A) \big] \Theta\ .
\end{equation}

As before let us establish the consistency of the theory with 
the standard gauged supergravity. Compared to the massless type IIA theory
the effect of the non-trivial $(p^A, q_A)$ NS-fluxes is the replacement
of ordinary derivatives by the covariant derivatives 
(\ref{NScd}), (\ref{NScda}) and the appearance of
the potential (\ref{VNS}). Otherwise the structure of the
theory is unchanged.
To show agreement with gauged supergravity we need to
demonstrate that with the gaugings (\ref{NScd}) and (\ref{NScda})
the potential (\ref{pot}) reduces to (\ref{NSpotE}).
The key point here is to notice that for the case at hand
(\ref{pot}) considerably simplifies in the sense
that the term which contains the Killing prepotentials vanishes. To see this
we first note that since only one vector field $A^0$ participates
in the gaugings (\ref{NScd}) and (\ref{NScda}), the only non-trivial components
of $P^x_I$ will be the ones for which $I=0$. Using (\ref{eqkpot}),
(\ref{ImN-1e}) and the fact that $X^0 = 1$ one immediately sees that
\begin{equation}
  \label{simp}
  \frac12 (\IM \N)^{-1\,00} + 4 e^K X^0 \bar X^0 = 0 \ .
\end{equation}
Inserting (\ref{simp}) into  (\ref{pot}) we arrive at 
\begin{equation}
  \label{Vint}
  V_E = 4 e^K h_{uv} k^u_0 k^v_0\ ,
\end{equation}
where we already used $k_i=0$. The Killing vectors $k^u_0$
can be read off from the covariant derivatives (\ref{NScd}) and (\ref{NScda})
to be 
\begin{equation}\label{KillNS}
  k_0^{\xi^B} = p^B \, , \quad k_0^{\tilde \xi_B} = q_B \, , 
  \quad k_0^a = 2 e + 
  (p^A \tilde\xi_A - q_A \xi^A)\ .
\end{equation}
Using the metric components of the charged scalars from (\ref{qktNS}), the 
evaluation of (\ref{Vint}) precisely results in the potential (\ref{VNS}) 
and thus 
establishes the consistency with gauged supergravity.

As we said before the difference for NS-fluxes is that the potential 
depends on the scalars in
hypermultiplets $(\phi,z^a,\xi^A,\tilde\xi_A)$ while the vector
multiplet scalars $t^i$ remain undetermined.
Furthermore,the only gauge field which appears in the covariant derivatives is the
graviphoton $A^0$ while all other $A^i$ do not participate in the gauging.

%
\section{Type IIB compactified on \CY  threefolds with
  background fluxes and mirror symmetry} 
\label{IIB/CY3}\setcounter{equation}{0}

So far we concentrated on type IIA theories compactified
on Calabi-Yau threefolds $Y_3$ and derived the four-dimensional
effective theory when non-trivial background fluxes are turned on.
Without fluxes these theories are equivalent to type IIB 
compactified on the mirror threefold $\tilde Y_3$. 
At the level
of the low energy effective action the precise map between
these compactifications was derived in refs.\  \cite{BC,FeS,BGHL}.
However, when background flux is turned on, the validity of perturbative 
and non-perturbative dualities become obscure and is not fully understood
at present \cite{HLS,CKKL,Janssen,BBRS,CV,CKL,HS2}. 
Before we discuss this issue in more detail let us
focus on type IIB compactification
with background fluxes. Without fluxes the effective theory
is given in \cite{BC,BGHL} while turning on three-form flux
was considered in refs.\ \cite{JM,TV,PM1,CKLT,GD}.
Here we do not redo the computation in detail
but focus on the situation where both electric and magnetic charges
$(e_A,m^A)$ are present. For this case the potential
has been computed in \cite{JM,TV} but the complete couplings of 
the two forms were not derived. (Ref.\ \cite{GD}
only considered the case $m^A=0$.)
The purpose of this section is to give the complete bosonic Lagrangian
including the couplings and mass terms of the NS two-form $B_2$
and to discuss the mirror symmetry to type IIA compactification
with fluxes as derived in the previous section. 
For simplicity we only turn on the background fluxes
of the R-R two-form $C_2$.

Let us start by recalling the structure of the ten-dimensional type IIB
supergravity. The NS-NS sector features the graviton, 
a two-form $\hat B_2$ and the dilaton $\hat\phi$, 
while the R-R sector contains a second scalar
$l$, a second two form $\hat C_2$ 
and a four form $\hat A_4$ with a self-dual field strength. 
The bosonic part of the low energy
effective action in ten dimensions reads \cite{JP}\footnote{In this action the
  self-duality condition on the  5-form field strength has not been imposed. A
  covariant action including the self-dual 4-form has been constructed in
  \cite{DLS}. The field equations for type IIB supergravity were originally
  derived in refs.\ \cite{schwarz}.} 
\begin{eqnarray}
  \label{Sten}
  S_{\textrm{IIB}} & = & \int e^{-2\hat \phi} \left(\frac12 \hat R
  \rule{0pt}{10pt}^*{\bf 1} + 2 d \hat \phi \awg d \hat \phi 
- \frac14 \hat H_3\awg \hat H_3
  \right) \nn \\ 
  & & - \frac12 \left(dl \awg dl + \hat F_3\awg \hat F_3\ +
  \frac12 \hat F_5 \awg \hat F_5 \right) - \frac12 \hat A_4\wg \hat H_3
  \wg d\hat C_2 \, ,
\end{eqnarray}
where the field strengths are defined as\footnote{%
Commonly one uses the convention $F_5' = d \hat A'_4 -\frac12 \hat C_2
  \wg d \hat B_2 + \frac12 \hat B_2 \wg d \hat C_2$
which is obtained from $F_5$ 
by the redefinition $A_4 = A'_4 - \frac12 B_2\wg C_2$.
Note that this redefinition
modifies the topological term only by a total derivative while the
self-duality constraint $\rule{0pt}{10pt}^*F_5 = F_5$ remains unchanged.}
\begin{equation}
  \label{IIBfs}
  \hat H_3 =  d\hat B_2\ ,\qquad
  \hat  F_3 = d \hat C_2 - l d \hat B_2 \, , \qquad 
  \hat F_5 = d \hat A_4  + \hat B_2 \wg d \hat C_2 \, .
\end{equation}


When compactified on a Calabi-Yau threefold the resulting low energy
spectrum contains as in type IIA 
$h_{1,1}$ K\"ahler deformations $v^i, i=1,\ldots,h_{1,1}$
of the Calabi-Yau metric and  
 $h_{1,2}$ complex deformations
${z}^{a}, a= 1,\ldots,h_{1,2}$ 
of the complex structure.
The doublet of two-forms  $\hat B_{2}, \ \hat C_2$ and 
the four-form $\hat A_4$ decompose according to 
\begin{eqnarray}
  \label{CYBred}
  \hat B_2 & = & B_2 + b^i\, \ox_i \ ,\qquad
  \hat C_2 =  C_2 +  c^i\, \ox_i\ ,\nn \\
  d \hat A_4 &=&  d D_2^i  \wedge \ox_i + F^A \ax_A - G_B \beta^B
  + d \rho_i \tilde \ox^i \ ,
 \end{eqnarray}
where as before $(\alpha_A, \beta^B)$
span a real $(2h_{2,1}+2)$-dimensional basis of $H^3(Y_3)$,
$\ox_i$ is a $h_{1,1}$-dimensional basis of $H^{1,1}(Y_3)$
and $\tilde \ox^i$ is the dual basis on $H^{2,2}(Y_3)$. The normalizations of
these basis are given in appendix \ref{CYM}.  
The self-duality condition $\rule{0pt}{10pt}^*F_5 = F_5$
implies that the $G_B$
are the dual magnetic field strengths  of $F^A$ while $\rho_i$ are the duals
of the tensors $D^i$. Together these fields 
 combine into a gravitational multiplet with bosonic components
$(g_{\mu\nu},A^0)$, a double-tensor multiplet
$(B_2, C_2,\phi,l)$, $h_{1,1}$ tensor multiplets  
$(D^i_{2},v^i, b^i, c^i)$ and $h_{1,2}$ vector multiplets
$(A^a,z^a)$.\footnote{%
We consider $A^A$ to be electric potentials, i.e. $F^A = d A^A$}
Dualizing the two-forms to scalars turns the tensor and double tensor
multiplets into 
hypermultiplets each containing four real scalars.
In this dual basis the low energy spectrum features  $h_{1,1}+1$
hypermultiplets and $h_{1,2}$ vector multiplets apart from the gravitational
multiplet.

Next we turn on background fluxes for the R-R three-form by
modifying $d \hat C_2$ 
\begin{equation}
  \label{fIIB}  
  d \hat C_2 \to d \hat C_2 + m^A \ax_A - e_A \bx^A \, ,
\end{equation}
where $e_I^A, m_I^A$ are the constant background fluxes. 
The field strength $\hat F_3$ turns into
\begin{eqnarray}
  \label{F3}
 \hat  F_3 = F_3 - l H_3 + (dc^i -ldb^i) \, \ox_i + m^A \ax_A - e_A \bx^A \, ,
\end{eqnarray}
where 
$F_3 = dC_2$ and $ H_3 =dB_2$ and
the only modification due to (\ref{fIIB}) coming from the term $F_3 \awg F_3$ 
is a potential 
\begin{equation}
  \label{potIIB}
  V  =  - \frac{1}{2} \big(e_A - m^C \Mc_{CA} \big) (\IM
  \Mc)^{-1AB} \big(e_B -  \Mc_{BD}m^D \big) \, .
\end{equation}
The matrix $\Mc$ is defined in (\ref{N_G}) and to obtain the above
expression we used (\ref{star}) and (\ref{A-N}).
Furthermore, $d \hat C_2$ appears in the definition of $\hat F_5$ (\ref{IIBfs})
which is modified according to
\begin{equation}
  \label{mF5}
  \hat{F}_5 = (dD_2^i + b^i dC_2 + B_2 \wg d c^i)\wedge \ox_i + \check F^A \ax_A
  - \check G_A \bx^A + (d \rho_i + b^j d c^k \K_{ijk}) \wg \tilde \ox^i \, , 
\end{equation}
where we defined
\begin{equation}
  \label{FtilB}
  \check F^A \equiv F^A + m^A B_2 \ , \qquad 
\check G_A \equiv G_A + e_A B_2 \, ,
\end{equation}
and used (\ref{oxstar}) to express the product of harmonic two forms as a
harmonic four form.
Finally, the topological term produces a Green-Schwarz type interaction
\begin{equation}
  \label{mcs}
  \dx \L_{\textrm{top}} = 
  - \frac12 (F^A e_A - G_A m^A) \wg B_2 \, .
\end{equation}

In the standard compactification of the type IIB theory the four dimensional
action is obtained after imposing the self-duality condition for $\hat
F_5$. This however, cannot be done directly since imposing $\hat F_5 =
\rule{0pt}{8pt}^* \hat F_5$ in (\ref{Sten}) the kinetic term for $\hat A_4$
vanishes identically and the same happens with the kinetic terms of the
fields which come 
from the reduction of $\hat A_4$. 
The correct 4 dimensional action is obtained by
adding appropriate Lagrange multipliers in order to impose the self-duality
condition. At the level of the reduced fields this condition can be obtained
from (\ref{mF5}) using the expressions for the Hodge duals of the harmonic
forms on a \CY 
threefold (\ref{oxstar}), (\ref{star}), (\ref{A-N}) and reads
\begin{equation}
  \label{Gtil}
  \check G_A =  \IM \Mc_{AB} \rule{0pt}{10pt}^*\check F^B  + \RE
  \Mc_{AB} \check F^B \, .
\end{equation}
Similarly one obtains
\begin{equation}
  \label{Ddx}
  d \rho_i + \K_{ijk} b^j dc^k = 4\K g_{ij} \rule{0pt}{8pt}^* (d D^j + b^j dC_2
  + B_2 \wg d c^i )\ .
\end{equation}
Strictly speaking this relation should involve field strengths as in
(\ref{Gtil}) rather than only exterior derivatives of some
potentials. However, since the fluxes play no role in this relation
we do not need to be  more precise here and instead rely to the 
standard \CY compactification of
type IIB theory. 
On the other hand, (\ref{Gtil}) does depend on the flux parameters
through $\check F$ and $\check G$ defined in (\ref{FtilB}) and reduces to the
usual formula (\ref{eom}) which relates the electric and magnetic field strengths
only for vanishing fluxes. To obtain the Lagrangian for the vector fields we
proceed as in the massless case \cite{GD}. First, treating $F^A$ and $G_A$
as independent fields, inserting (\ref{mF5}) into (\ref{Sten}) and taking into 
account (\ref{mcs}) we obtain 
\begin{eqnarray}
  \label{FG}
  \L(F,G) & = &  \frac14 (\IM \Mc)^{-1AB}(\check G_A - \check F^C \Mc_{AC})
  \wg (\,\rule{0pt}{10pt}^* \check G_B - \rule{0pt}{10pt}^* \check F^D  \bar
  \Mc_{BD}) \nn \\
  & & - \frac12 (F^A e_A - G_A m^A) \wg B_2\ .
\end{eqnarray}
Furthermore, to impose (\ref{Gtil}) as the equation of motion for $G$ we have
to add to (\ref{FG}) the term $\frac12 F^A \wg G_A$. Eliminating $G_A$
using (\ref{Gtil}) one is left with the following Lagrangian for $F^A$ 
\begin{eqnarray}
  \label{LF}
  \L(F) & = & \frac12 F^A \wg G_A - \frac12 (F^A e_A - G_A m^A) \wg B_2  \\
  & = & \frac12 \IM \Mc_{AB} \check F^A \awg \check F^B + \frac12 \RE \Mc_{AB}
  \check F^A \check F^B - \frac12 (F^A e_A + \check F^A e_A) \wg B_2 \, . \nn
\end{eqnarray}

The scalar sector is not modified by the introduction of fluxes in
(\ref{fIIB}) and so we can use the results in the literature 
\cite{BC,FeS,BGHL}
to obtain the following 4 dimensional action 
\begin{eqnarray}
  \label{SIIB4}
  S & = & \int e^{-2\phi} \Big( \frac12 R ^\ast \! {\bf 1} + 2d \phi
  \awedge d \phi - \frac14 H_3 \awedge H_3 - g_{ab} dz^a \awg dz^b 
  - g_{ij} dt^i \awg d{\bar t}^j \Big)   \nn \\
  && + \frac{1}{2}\left(\IM \N^{-1} \right)^{IJ}  
  \Big[ d\tilde\xi_I + \N_{IK} d\xi^K \Big] 
  \awg  \Big[ d\tilde\xi_J + \bar \N_{JL} d\xi^L \Big]  \, .\nn\\
  & & +\frac12 
  H_3 \wg (\tilde\xi_A d \xi^A - \xi_A d \tilde\xi_A )
  + \frac12 \IM \Mc_{AB} \check F^A \awg \check F^B 
  + \frac12 \RE \Mc_{AB} \check F^A \wg \check F^B \nn \\
  & & -\frac12 B_2 \wg  (\check F^A + dA^A) e_A  - V \ ,
\end{eqnarray}
where $V$ is given in (\ref{potIIB}) and the scalars $\xi^I$ and
$\tilde \xi_I$ are functions of $l,c^i,\rho_i$ and the dual of $C_2$
specified by the
mirror map \cite{BGHL}.

Mirror symmetry at the level of the low energy effective actions is now
obvious since the (\ref{AIB}) and (\ref{SIIB4}) have the same form. The only
difference is the range of indices and the fact that the matrices $\N$ and
$\Mc$ are interchanged which just expresses the mirror symmetry
$h_{1,1} \leftrightarrow h_{1,2}$. Moreover the potentials (\ref{potIIB}) and
(\ref{Apot}) are the same and the flux parameters $(m^A, e_A)$ in type IIB
case are the mirror partners of $(m^I, e_I)$ in type IIA. 
Due to the appearance of $\check F$ in (\ref{SIIB4}) $B_2$ is also massive in 
IIB -- a fact which has not
been stressed previously. Finally,
the two extra parameters in type IIA, the one coming from the mass
$m$ of the massive type IIA supergravity and the other coming from the
dualization of the 4 dimensional $C_3$ to a constant $e_0$, are crucial in
order to make mirror symmetry work when R-R fluxes are turned on.

A similar analysis can be performed for the NS three-form.
Apart from the factor of the dilaton one obtains the same result as 
computed in refs.\ \cite{JM,TV,GD}. However, in this case 
there is no obvious mirror dual set of fluxes in type IIA.
This has led to the proposition \cite{CV} that in type IIA the
holomorphic three-form $\Omega$ ceases to be holomorphic and effectively 
an NS two- and four-form is induced. It would be interesting to make this 
proposal
more explicit at the level of the effective action.
Similarly there is no obvious mirror dual in type IIB of the 
NS-fluxes $(p^A, q_B)$ discussed in section~2.4.

\section{Conclusions}
\label{conc}\setcounter{equation}{0}

In this paper we studied the compactification of type II theories on \CY
threefolds in the presence of background fluxes. First we concentrated on the
massive type IIA theory and considered non-trivial background
values for the R-R fields $\hat F_2, \hat F_4$. 
These fluxes are in one
to one correspondence with harmonic $(1,1)$- and $(2,2)$-forms on the \CY
manifold and therefore induce $2 h_{1,1}$ flux parameters into the effective 
action.  Furthermore,
dualizing the three-form $C_3$ in the $d=4$ effective action
to a constant provides an additional parameter. 
Together with the (ten-dimensional) mass of massive type IIA theory
one finds a total of $2 h_{1,1}+2$ flux parameters which
naturally combine into symplectic vector of $Sp(2 h_{1,1}+2)$. 
Furthermore, we showed 
that in fact the entire theory is invariant under $Sp(2 h_{1,1}+2)$
rotations due to the presence of a massive two-form $B_2$ which
couples to both electric and magnetic field strengths. 
In our analysis we just followed the compactification
of  type IIA supergravity but there is obviously a more general story 
to be discovered here. It would be interesting to construct the 
most general gauged supergravity including massive two-forms
and investigate the conditions for the symplectic
$Sp(2 h_{1,1}+2)$ invariance of the theory entirely within supergravity
and without any particular reference to flux backgrounds of string theory. 

For vanishing magnetic fluxes $m^I =0$ the two-form 
$B_2$ is massless and can be dualized to a scalar $a$. 
This dual basis is a standard
gauged $N=2$ supergravity with a potential (\ref{ApotE}) 
depending on the scalars in the vector multiplets.
Only $a$ is charged but with respect to linear
combination of all $h_{1,1}+1$ gauge fields.
The form of the potential coincides with the potential found 
in \cite{LM} for the case of
the heterotic string compactified on $T^2 \times K3$ 
when a set of very specific fluxes along $K3$
are turned on. In particular, in the heterotic case more than one
scalar is generically charged and thus the fluxes have to be chosen 
to point in the direction of $a$ in order to find agreement
with the type IIA case \cite{CKKL}.

The perturbative duality of type IIA compactified
on $Y_3$ to type IIB compactified on the
mirror threefold $\tilde Y_3$ was established at the level
of the effective Lagrangian for R-R fluxes turned on in both theories.
For non-vanishing $m^I$ also the type IIB effective theory
features a massive $B_2$. Furthermore, 
the validity of the duality crucially depends on including
the ten-dimensional mass parameter $m^0$ and the dual constant $e_0$
of three-form $C_3$. 

The situation for NS-fluxes remains murky. We derived the 
effective theory of type IIA in the presence of NS
three-form flux $H_3$. In this case a standard gauged
$N=2$ supergravity is found but contrary to R-R flux backgrounds 
the potential 
depends non-trivially on the scalars in the hypermultiplet.
Furthermore,
the hyper-scalars are charged only with respect to the graviphoton
but not with respect to any of the other $h_{1,1}$ gauge fields.

In the type IIB compactification we found no obvious set of NS mirror
fluxes and leave this puzzle for further studies.
As suggested in \cite{CV} it might be related to the fact
that the Calabi-Yau geometry is deformed and the precise
nature of the deformation has to be taken into account
in more detail.

We also did not address the 
issue of holomorphic superpotentials in relation
with spontaneous $N=2\to N=1$ supersymmetry breaking \cite{TV}
which we leave to a separate publication \cite{GL}.


\vspace{1cm}
\appendix
\noindent
{\Large {\bf Appendix}}
\renewcommand{\theequation}{\Alph{section}.\arabic{equation}}

\setcounter{equation}{0}\setcounter{section}{0}
\section{Notations and conventions}
\label{conv}
In this appendix we assemble the conventions used throughout the paper.
\begin{itemize}
\item The space-time metric has signature $(-, +, +, \ldots)$.

\item The components of a differential $p$-form are defined as follows
  \begin{equation}
    A_p = \frac{1}{p !} A_{\mu_1 \ldots \mu_p} d x^{\mu_1} \wg \ldots 
\wg d x^{\mu_p} \, .
  \end{equation}

\item A hat on a $p$-form, e.g.\  $\hat{A_p}$ denotes 
differential forms in $d=10$. $p$-forms without the hat are
four-dimensional quantities.

\item The Hodge operation $\rule{0pt}{10pt}^*$ is defined in such a way that 
  \begin{equation}
    \label{Hstar}
    d A_p \awg d A_p = \frac{\sqrt{-g}}{p!}\, (dA)_{\mu_1 \ldots \mu_{p+1}}
    (dA)^{\mu_1 \ldots \mu_{p+1}}  d^d x
  \end{equation}
reproduces the correct kinetic term for a $p$-form in $d$ 
space-time dimensions. In particular we denote
$\rule{0pt}{8pt}^* {\bf 1} = \sqrt{-g}
\, d^d x$.

\item After compactification the Hodge operation splits into a Hodge-star on the
  four-dimensional space and another one acting on the internal \CY space. For
  example, in the expansion of a $p$ form one encounters terms like $\hat
  A_p = \cdots + A_{p-k} \ox_{k} + \cdots$, where $\ox_k$ is some harmonic $k$
  form on the internal space. The Hodge dual is given by 
  \begin{equation}
    \rule{0pt}{10pt}^* \! \! \hat A_p = \cdots + (-1)^{k(p-k)} \,
    \rule{0pt}{10pt}^* \!\!A_{p-k} \rule{0pt}{10pt}^* \ox_k + \cdots \, ,
  \end{equation}
where the first $\rule{0pt}{10pt}^*$ on the RHS acts only in space-time
while the second acts only in the internal space. The $(-1)^{k(p-k)}$ assures
that the kinetic term of $\hat A_p$ produces 
\begin{equation}
  \int_{Y_3} \hat A_p \awg \hat A_p = \cdots + A_{p-k} \awg A_{p-k} \int_{Y_3}
  \ox_k \awg \ox_k + \cdots .
\end{equation}
\item The indices $i,j,k, \ldots $ label harmonic $(1,1)$ and $(2,2)$ forms
  on the \CY threefold and run from $1$ to $h_{1,1}$; 
the indices $I, J, \ldots$
  label the vector fields in type IIA compactifications and include the zero
  $I=0,1, \ldots, h_{1,1}$. The  indices $a, \ b \ldots$ run
  from $1$ to $h_{1,2}$ and label $(1,2)$-forms on $Y_3$. The indices 
$A, \ B \ldots$ include the zero and label all three-forms including
the $(3,0)$-form, i.e.\ $A=0,1, \ldots, h_{1,2}$. $A, B, \ldots $ also
label vector fields in type IIB compactifications.

\end{itemize}

\section{$N=2$ supergravity in $d=4$}
\setcounter{equation}{0}
\label{sg}

The purpose of this appendix is to give a short review about $N=2$ 
supergravity in $d=4$ \cite{wp,bw,N=2,CDF}. 
A generic spectrum contains the gravitational multiplet,
$n_V$ vector multiplets and $n_H$ hypermultiplets.
The vector multiplets contain $n_V$ complex scalars
$t^i, i=1,\ldots,n_V$ while the hypermultiplets
contain $4n_H$ real scalars $q^u, u=1,\ldots, 4n_H$.\footnote{%
There also exist $N=2$  tensor multiplets which contain
a two-form $B_2$ and three real scalars as bosonic components.
Upon dualizing $B_2$ to a scalar the tensor multiplet
can be treated as an additional hypermultiplet. With this in mind  
 we do not discuss tensor multiplets
in this appendix.}
Due to supersymmetry the scalar manifold factorizes
\begin{equation}\label{modulispace}
\cM=\cM_V \otimes \cM_H \ ,
\end{equation}
where the component
$\cM_V$ is a  special K\"ahler manifold
spanned by the scalars $t^i$ while
$\cM_H$ is a quaternionic manifold spanned by the scalars $q^u$.

A special K\"ahler manifold is  a K\"ahler manifold  whose geometry 
obeys  an additional constraint \cite{wp}.
This constraint states that the 
 K\"ahler potential $K$ is not an arbitrary real function 
but 
determined in terms of a holomorphic prepotential $\F$
according to 
\begin{equation}\label{Kspecial}
K=-\ln \Big(i \bar{X}^{I} (\bar t) \F_{I}(X)
- i X^{I}  (t)\bar{\F}_{I}(\bar{X})\Big) \ ,
\end{equation}
where $X^{I}, I=0,\ldots, n_V$ are $(n_V+1)$
holomorphic functions of the 
$t^i$.
$\F_{I}$ abbreviates the derivative, i.e.
$\F_{I}\equiv \frac{\partial \F(X)}{\partial X^{I}} $
and 
$\F(X)$ is  a homogeneous function of $X^{I}$ of degree $2$, i.e.\
$X^{I} \F_{I}=2 \F$.

The $4n_H$ scalars $q^u, u=1,\ldots,4n_H$ in the hypermultiplets are
coordinates on a quaternionic manifold \cite{bw}.
This implies the existence of three almost complex
structures $(J^x)^w_v, x=1,2,3$ which satisfy the
quaternionic algebra
\begin{equation}\label{Jx}
J^x J^y = -\delta^{xy} + i \epsilon^{xyz} J^z\ .
\end{equation}
Associated with the complex structures is a
triplet of K\"ahler forms
\begin{equation}\label{Kx}
K^x_{uv} = h_{uw} (J^x)^w_v\ ,
\end{equation}
where $h_{uw}$ is the quaternionic metric.
The holonomy group of a quaternionic manifold
is $Sp(2)\times Sp(2n_h)$ and
$K^x$ is identified with the field strength of the
$Sp(2)\sim SU(2)$ connection $\omega^x$, i.e.
\begin{equation}\label{Kdef}
K^x \ = \
d\omega^x +\frac12\,\epsilon^{xyz}\omega^y\wedge \omega^z~.
\end{equation}

The bosonic part of the (ungauged) $N=2$ action is given by
\begin{equation}
\label{asg}
  S =  \int \Big[ \frac12 R ^* {\bf 1} - g_{i\bar\jmath} dt^i \awg d
  {\bar t}^j - h_{uv} dq^u \awg dq ^v 
  + \frac{1}{2}\, \IM \N_{IJ} F^I\awg F^{J}
  + \frac{1}{2} \, \RE \N_{IJ} F^I \wg F^J
  \Big] \ ,
\end{equation}
where $g_{i\bar\jmath}=\partial_i\partial_{\bar\jmath} K,\ F^I = dA^I$
($F^0$ denotes the field strength of the graviphoton)
and the gauge coupling functions are given by
\begin{equation}
\label{Ndef}
{\cal N}_{IJ} = \bar \F_{IJ} +2i\ \frac{\mbox{Im} \F_{IK}\mbox{Im} \F_{JL} X^K
 X^L}{\mbox{Im} \F_{LK}  X^K X^L} \ . 
\end{equation}

The equations of motion of the action (\ref{asg}) are invariant under
generalized electric-magnetic duality transformations. 
{}From (\ref{asg}) one derives the equations of motions
\begin{equation}
  \label{eom}
\frac{\partial \L}{\partial A^I} = 
d {G}_I = 0 \ , \qquad 
{G}_I = \RE \N_{IJ} {F}^J 
+ \IM \N_{IJ} \rule{0pt}{10pt}^*{F}^J \, ,
\end{equation}
while the Bianchi identities read
\begin{equation}
d {F}^I = 0\ .
\end{equation}
These equations are invariant under the generalized duality rotations\footnote{
This is often stated in terms of the self-dual and anti-self-dual part of the
field strength $F^{\pm J}$ and the dual quantities
$G^+_{I}\equiv{\cal N}_{IJ}F^{+J}\,,\
G^-_{I} \equiv\bar{\cal N}_{IJ}F^{-J}$.}
\begin{eqnarray}\label{FGdual}
F^{I}&\to&
U^I{}_J\, F^{J}+Z^{IJ}\,G_{ J}\ ,\nn\\
G_I&\to& V_I{}^J\,G_{J}+W_{IJ}\,F^{J}\ , 
\end{eqnarray}
where $U$, $V$, $W$ and $Z$ are constant, real,  $(n_V+1)\times(n_V+1)$
matrices which obey
\begin{eqnarray}
  \label{spc2}
  U^{\rm T} V- W^{\rm T} Z &=& V^{\rm T}U - Z^{\rm T}W =
  {\bf 1}\, ,\nn\\
  U^{\rm T}W = W^{\rm T}U\,, && \quad Z^{\rm T}V= V^{\rm T}Z\ .
\end{eqnarray}
Together they form the $(2n_V+2)\times(2n_V+2)$ symplectic matrix 
\begin{equation}
  \label{uvzwg}
  {\cal O}\ = \left(
    \begin{array}{cc} 
      U & Z \\
      [1mm] W & V 
    \end{array} 
  \right) \, , \qquad {\cal O}\in Sp(2n_V+2)\ .
\end{equation}
$(F^I,G_I)$ form
a $(2n_V+2)$ symplectic vector; $(X^I,\F_I)$ enjoy the same
transformations properties and also transforms as a
symplectic vector under (\ref{FGdual}). Clearly, the K\"ahler  potential
(\ref{Kspecial}) is invariant under this symplectic transformation.
Finally, the matrix $\N$ transforms according to 
\begin{equation}
  \label{nchange}
  \N \to (V \N+ W) \,(U+ Z \N)^{-1} \,.
\end{equation}

Let us now turn to gauged $N=2$ supergravity \cite{N=2}.
One can gauge 
the isometries on the scalar manifold ${\cal M}$.
Such isometries are generated by the
Killing vectors $k_I^u(q),\, k_I^i(t)$
\begin{equation}\label{kdef}
\delta q^u \ = \ \Lambda^I k_I^u(q) \  ,\qquad
\delta t^i \ = \ \Lambda^I k_I^i(t) \  .
\end{equation}
$k_I^u(q),\, k_I^i(t)$ satisfy the Killing equations
which in $N=2$ supergravity can be solved in terms of
four Killing prepotentials $(P_I,P_I^x)$.
The Killing vectors on $\cM_V$ are holomorphic
and obey
\begin{equation}
k_I^i(t) = g^{i\bar j} \partial_{\bar j} P_I\ ,
\end{equation}
while 
the Killing vectors on $\cM_H$ are determined by a
triplet of Killing prepotentials $P_I^{x}(q)\,$
via
\begin{equation}\label{killingpre}
k^u_I\,K_{uv}^x \ = \  -D_v P_{I}^x \equiv
- (\partial_v P_{I}^x
+ \epsilon^{xyz} \omega_{v}^y P_I^z)\ .
\end{equation}
Gauging the isometries (\ref{kdef}) requires the replacement of 
ordinary derivatives by covariant derivatives in the action
\begin{equation}\label{gaugeco}
\partial_\mu q^u \to {D}_\mu q^u = \partial_\mu q^u - k_I^u A_\mu^I\ ,
\qquad
\partial_\mu  t^i \to 
{D}_\mu  t^i = \partial_\mu  t^i - k_I^i A_\mu^I\ .
\end{equation}
Furthermore the potential 
\begin{eqnarray}
  \label{pot}
  V_E  & = & \ e^K \Big[ X^I\bar X^J (g_{\bar \imath j}\, k_I^{\bar\imath} k_J^j +
  4h_{uv}\,k^u_Ik^v_J) + g^{i\bar \jmath} D_i X^I D_{\bar j} 
  \bar X^J P_I^x P_J^x - 3 X^I \bar X^J P_I^x P_J^x \Big] \nn\\
  & = & e^K X^I \bar X^J (g_{\bar \imath j} \, k_I^{\bar\imath} k_J^j +
  4h_{uv}\, k^u_I k^v_J) -  \Big[ \frac12 (\IM \N)^{-1 IJ} + 4 e^K X^I \bar X^J \Big]
 P_I^x P_J^x \ , \hspace{1cm}
\end{eqnarray}
has to be added to the action in order to preserve supersymmetry.
The bosonic part of the action of gauged $N=2$ supergravity is 
then given by 
\begin{equation}
  \label{agsg}
  S =  \int  \frac12 R ^* {\bf 1} - g_{i\bar\jmath} Dt^i \awg D
  {\bar t}^j - h_{uv} Dq^u \awg Dq ^v 
  + \frac{1}{2}\, \IM \N_{IJ} F^I\awg F^{J}
  + \frac{1}{2} \, \RE\N_{IJ} F^I \wg F^J -V_E \ .
\end{equation}
The symplectic invariance of the ungauged theory is generically
broken since the action now explicitly depends on the 
gauge potentials $A^I$ through the covariant derivatives $Dt^i, Dq^u$.

\section{The moduli space of \CY threefolds}
\label{CYM}\setcounter{equation}{0}

The moduli space of Calabi-Yau threefolds $Y_3$ splits into the space of 
K\"ahler deformations of the Calabi-Yau metric and deformations 
of the complex structure. The K\"ahler deformations are harmonic
$(1,1)$-forms and thus are elements of $H^{1,1}(Y_3)$. 
The complex structure deformations
are harmonic $(1,2)$-forms and thus are elements of
$H^{1,2}(Y_3)$. In string theory a two-form $B_2$ always appears
together with the (space-time) metric in the NS-NS sector. Its
compactification on a \CY threefold produce $h_{1,1}$ additional scalars
which together with the K\"ahler class deformations form the the
``complexified K\"ahler cone" $\cM_{1,1}.$
The moduli space $\cM$ of Calabi-Yau manifolds 
is be a direct
product of $\cM_{1,1}$ 
and the space $\cM_{1,2}$ spanned by the complex structure deformations
\begin{equation}
\cM = \cM_{1,1} \times  \cM_{1,2}\ .
\end{equation}
$\cM_{1,1}$ and $\cM_{1,2}$ are both
special K\"ahler manifolds and $\cM_{1,1}$ describes the vector multiplet
sector in type IIA theories while $\cM_{1,2}$ characterizes the vector
multiplet sector in type IIB case. In this appendix we briefly summarize these
geometries following refs.\  \cite{CdO,CDF,Suz,AS2}.

\subsection{The complexified K\"ahler cone}
\label{CKC}

The K\"ahler class deformations 
together with the zero modes of 
$B_2$ are harmonic $(1,1)$-forms on $Y_3$. Hence both the K\"ahler form $J$
and $B_2$ can be expanded in a basis $\omega_i$  of $H^{1,1}(Y_3)$
\begin{equation}
 B_2 +i J\ =\ (b^j + i v^j) \omega_j\ \equiv\  t^j \omega_j \ , \qquad 
j=1, \ldots, h_{1,1}\ .
\end{equation}
It is useful to define the  following quantities:
\begin{eqnarray}
  \label{K}
  \K & = & \frac16 \int_{Y_3} J \wg J \wg J  \, , \quad  \K_i = \int_{Y_3}
  \ox_i \wg J \wg J \ , \\ 
  \K_{ij} & = & \int_{Y_3} \ox^i \wg \ox ^j \wg J \ , \quad \K_{ijk} =
  \int_{Y_3} \ox^i \wg \ox^j \wg \ox^k  \ .\nn 
\end{eqnarray}
Note that we have introduced the factor $\frac16$ in the definition of  $\K$
so that it is precisely the volume of $Y_3$. 
The metric on the complexified K\"ahler cone  $\cM_{1,1}$ is K\"ahler, i.e.\
$g_{i j} = \del_i \bar \del_j K$ and 
given by  \cite{CdO, BCF} 
\begin{equation}
  \label{gH11}
  g_{i j}  = \frac{1}{4 \K} \int_{Y_3} \ox_i \awg \ox_j = -\frac14
  \left(\frac{\K_{i \bar \jmath}}{\K} -  \frac14 \, \frac{\K_i \K_{\bar
  \jmath}}{\K^2}\right) 
  = - \del_i \del_j \big(- \ln{8 \, \K} \big)  =\del_i \bar \del_j
  \big(- \ln{8 \, \K} \big) \ .
\end{equation}
Furthermore, the K\"ahler potential $K$
is determined in terms of a holomorphic prepotential $\F$ via
\begin{equation}
  \label{eqkpot}
  e^{-K} = 8 \, \K = i \left( {\bar X}^I \F_I - X^I {\bar \F}_I \right)
  \ , \qquad  \F_I \equiv \del_I \F \ , \qquad I=0, \ldots, h_{1,1}\ ,
\end{equation}
where 
\begin{equation}
  \label{prepot}
  \F = - \frac{1}{3!} \, \frac{\K_{ijk} X^i X^j X^k}{X^0}\ .
\end{equation}
The  (complex) K\"ahler class deformations $t^i$ are the so called
special coordinates related to the $X^I$ via
$X^I = (1, t^i)$. $(X^I, F_I)$ transforms as a symplectic vector under
(\ref{uvzwg}) and $K$ is a symplectic invariant.

The scalar manifold $\cM_{1,1}$ describes the moduli space of the vector
multiplets in the low energy effective action of type IIA supergravity 
compactified on a Calabi-Yau threefold. Therefore the matrix $\N$ 
defined in (\ref{Ndef}) plays
the role of generalized gauge couplings. Inserting (\ref{prepot})
into (\ref{Ndef}) it is straightforward to derive 
\begin{eqnarray}
  \label{eq:N}
  \RE \N_{00} = - \frac13 \K_{ijk} b^i b^j b^k \, , & & \IM \N_{00} = -
  \K + \left(\K_{ij} - \frac14 \, \frac{\K_i \K_j}{\K} \right) b^i
  b^j\ ,
  \nn \\
  \RE \N_{i0} = \frac12 \K_{ijk} b^j b^k \ , & & \IM \N_{i0} = -
  \left(\K_{ij} - \frac14 \, \frac{\K_i \K_j}{\K} \right) b^j \ ,\\
  \RE \N_{ij} = - \K_{ijk} b^k \ , & & \IM \N_{ij} = \left(\K_{ij} - \frac14
  \, \frac{\K_i \K_j}{\K} \right) \ .\nn 
\end{eqnarray}

In the main text we encounter the inverse matrices
$g^{ij}$ and $ (\IM \N)^{-1IJ} $. These can be expressed in terms of
harmonic $(2,2)$ forms. On a \CY threefold $H^{2,2} (Y_3)$ is dual to
$H^{1,1} (Y_3)$ and it is useful to introduce the dual basis $\tilde \ox^i$
normalized by
\begin{equation}
  \label{normH2}
  \int_{Y_3} \ox_i \wg \tilde \ox^j\ =\ \dx_i^j\, .
\end{equation}
With this normalization the following relations hold
\begin{equation}
  \label{oxstar}
  g^{ij} = 4 \K \int_{Y_3} \tilde \ox^i \awg \tilde \ox^j \, , \quad
  \rule{0pt}{8pt}^* \ox_i = 4 \K g_{ij} \tilde \ox^j \, , \quad 
  \rule{0pt}{8pt}^* \tilde \ox^i = \frac{1}{4 \K} g^{ij} \ox_j \, ,
\quad \ox_i \wg \ox_j \sim \K_{ijk} \tilde \ox^k \, ,
\end{equation}
where 
the symbol $\sim$
denotes the fact that the quantities are in the same cohomology class.
Introducing $\K^{ij}$ via
\begin{equation}
  \label{Kinv}
  \K^{ij} \K_{jk} = \dx^i_k \, ,
\end{equation}
one derives 
\begin{equation}
  \label{ImN-1e}
  \left(\IM \N\right)^{-1} = \left(
    \begin{array}{cc}
      - \frac{1}{\K} & - \frac{b^i}{\K} \\
      & \\
      - \frac{b^i}{\K}\quad & \K^{ij} - \frac{b^i b^j}{\K} - \frac{v^i
      v^j}{2 \, \K} 
    \end{array} \right) \, .
\end{equation}
Using (\ref{Kinv}) one can also write an explicit formula for the inverse
metric $g^{ij}$
\begin{equation}
  \label{Minv}
  g^{ij} = - 4 \K \left(\K^{ij} - \frac{v^i v^j}{2 \, \K} \right) \, .
\end{equation}

\subsection{The special geometry of $H^3$}
\label{SGH3}

The complex structure deformations of a threefold 
parameterize $H^{1,2}(Y_3)$. However, it turns out be convenient to discuss
the entire 
$H^{3}(Y_3) = H^{3,0}\oplus H^{2,1}\oplus  H^{1,2}\oplus H^{0,3}$
including $H^{3,0}$ and $H^{0,3}$ which have only
the holomorphic $(3,0)$-form $\Omega$ and the complex conjugate 
$(0,3)$-form $\bar\Omega$ as elements. One commonly
chooses a real basis 
$(\ax_A, \bx^B)$ on $H^3$ which obeys
\begin{eqnarray}
  \label{norm}
  \int_{Y_3} \ax_A \wg \bx^B & =& \dx_A^B\ =\ 
-\int_{Y_3} \bx^B \wg \ax_A \ , \qquad A,B = 0,\ldots, h_{1,2}\ ,\nn\\
  \int_{Y_3} \ax_A \wg \ax_B &=& \int_{Y_3} \bx^A \wg \bx^B\ =\ 0\ .
\end{eqnarray}
Note that these relations are invariant under symplectic rotations
\begin{equation}
  \label{srot}
  \left(
  \begin{array}{c}
    \bx \\
    \ax \\
  \end{array} \right) \to \left(
  \begin{array}{cc}
    U & Z \\
    W & V \\
  \end{array} \right) \ \left(
  \begin{array}{c}
    \bx \\
    \ax \\
  \end{array} \right), 
\end{equation}
where the matrices $U, \ V, \ W, \ Z$ satisfy (\ref{spc2}).
The unique holomorphic $(3,0)$ $\Ox$ can be expanded in terms
of this basis according to 
\begin{equation}
  \Ox = Z^A \ax_A - \G_A \bx^A\ ,
\end{equation}
where  $Z^A,\G_A$ are  the periods of $\Ox$  defined as
\begin{equation}
  \label{csm}
  Z^A\ =\ \int_{Y_3} \Ox \wg \bx^A \ ,  \qquad 
\G_A\ =\ \int_{Y_3} \Ox\wg \ax_A\ .
\end{equation}
$\Ox$ is invariant under (\ref{srot}) and hence $(Z^A, \G_A)$ transforms as a
symplectic vector.
The $\G_A$ are functions of $Z^A$ and determined in terms of a
homogenous function of degree two $\G(Z)$ as
\begin{equation}
  \label{defG}
  \G_A = \frac{\partial\G}{\partial Z^A} \equiv \partial_A \G\ .
\end{equation}
Furthermore, $\Ox$ is homogenous of degree one in $Z$, i.e.\ 
$\Ox = Z^A \del_A \Ox$ with
\begin{eqnarray}
\del_A \Ox  =
  \ax_A - \G_{AB} \bx^B \ .
\end{eqnarray}

The deformations of the complex structure $z^a, a = 1,\ldots,h_{1,2}$
which reside in $H^{1,2}(Y_3)$ are related to the coordinates $Z^A$
via $z^a = Z^a/Z^0$ or in other words
one can choose $Z^A = (1,z^a)$.
The metric $g_{a\bar b}$ on the space of complex structure deformations 
$\cM_{1,2}$ is K\"ahler
$g_{a\bar b} = \del_a \bar\del_{\bar b} K$ with the K\"ahler potential $K$
given by 
\begin{equation}
  \label{kpot}
  K = - \ln{\, i \!\int_{Y_3} \Ox \wg \bar \Ox} = -\ln{\, i
  \left({\bar Z}^A \G_A - Z^A {\bar \G}_A \right)} \ .
\end{equation}
As we see $K$ is determined in terms of the holomorphic prepotential $\G(Z)$
and hence $\cM_{1,2}$ is a 
special K\"ahler manifold. 
Note that $K$ is symplectically invariant.

Finally, let us discuss the action of the Hodge $\rule{0pt}{10pt}^*$
on the basis (\ref{norm}). 
$^* \ax_A$ and $^* \bx^B$ are both three-forms  again so that they can be expanded 
in terms of $\ax$ and $\bx$ according to 
\begin{equation}
  \label{star}
  ^* \ax_A =  {A_A}^B \, \ax_B + B_{AB} \, \bx^B \ , \qquad
  ^* \bx_A = C^{AB} \, \ax_B + {D^A}_B \, \bx^B\ .
\end{equation}
Using (\ref{norm}) one derives 
\begin{eqnarray}
  \label{ABC}
  B_{AB} = \int_{Y_3} \ax_A \awg \ax_B & = & \int_{Y_3} \ax_B \awg \ax_A 
= B_{BA}\ , \nn
  \\
  C^{AB} = - \int_{Y_3} \bx^A \awg \bx^B & = & - \int_{Y_3} \bx^B \awg \bx^A = C^{BA}\ ,
  \nn \\
  {A_A}^B = - \int_{Y_3} \bx^B \awg \ax_A & = & - \int_{Y_3} \ax_A \awg \bx^B = -
  {D^B}_A \ .
\end{eqnarray}
Furthermore, the matrices $A, \ B, \ C$ can be determined in terms of a matrix
 $\Mc$ \cite{CDF,Suz}
\begin{eqnarray}
  \label{A-N}
  A & = & \left(\RE \Mc \right) \left(\IM \Mc \right)^{-1}\ , \nn \\
  B & = & - \left(\IM \Mc \right) - \left(\RE \Mc \right) \left(\IM
  \Mc\right)^{-1} \left(\RE \Mc \right)\ , \nn \\
  C & = & \left(\IM \Mc \right)^{-1}\ ,
\end{eqnarray}
where
\begin{equation}
  \label{N_G}
  \Mc_{AB} = {\bar \G}_{AB} + 2i \, \frac{(\IM \G)_{AC} Z^C (\IM \G)_{BD}
  Z^D}{Z^C (\IM \G)_{CD} Z^D}\ \ .
\end{equation}
$\Mc_{AB}$ determines the gauge couplings in type IIB compactifications
on $Y_3$.


\section{Massless type IIA supergravity compactified on \CY
threefolds without fluxes}
\label{IIA/CY3}\setcounter{equation}{0}

In this section we recall the compactification of massless type IIA
supergravity on a \CY threefold first performed in ref.\ \cite{BCF}. 
The purpose of this appendix is twofold. One the one hand we need
to redo the computation in order to fix
the notation and convention for the case studied
in section~\ref{RRf} where fluxes are turned on. On the other hand
the detailed dualization of the three-form $C_3$ to our
knowledge has not been presented previously. This is of importance
for our analysis in section~\ref{RRf} as the three-form $C_3$ turns out to be dual
to a constant $e_0$ which nicely combines with other flux parameters
to build symplectic invariant combinations.\footnote{In ref.\ \cite{BCF}
the case  $e_0=0$ was considered.}
Furthermore, as we will show $e_0$ is the charge
of the scalar $a$ which is dual to $B_2$ and a potential consistent
with the standard $N=2$ gauged supergravity is induced.

Let us start from the ten-dimensional action of massless type IIA supergravity.
It features the graviton, a two form  $\hat B_2$ and the dilaton $\hat\phi$ 
in the NS-NS sector,  a
vector field $\hat A_1$ and a three-form $\hat C_3$ in the R-R-sector and
reads 
\begin{equation}
  \label{S10'}
  S = \int \,e^{-2\hat\phi} \big( \frac12 \hat R ^\ast\! {\bf 1} + 2
  d \hat\phi \awedge d \hat\phi - \frac12 \hat H_3 \awedge \hat H_3 \big)  
  - \frac12  \, \big(
  \hat F_2 \awedge \hat F_2  + \hat F_4 \awedge  \hat F_4 \big)
  + \mathcal {L}_{top} \, ,
\end{equation}
where 
\begin{eqnarray}
\hat  F_4 &=& d\hat C_3 - d\hat A_1 \wg \hat B_2\ , \qquad \hat F_2 = d\hat A_1\ ,
\qquad \hat H_3 = d \hat B_2 \ ,\nn\\
\mathcal {L}_{top} &=& - \frac12\Big[
\hat B_2 \wg d\hat C_3 \wg d\hat C_3 - (\hat B_2)^2 \wg d\hat C_3 \wg
  d\hat A_1 + \frac13 (\hat B_2)^3\wg d\hat A_1 \wg d \hat A_1\Big]\ .
\end{eqnarray}
We have chosen the conventions such that (\ref{S10'}) is the $m\to 0$ limit
of the action (\ref{mIIA}).\footnote{There is an ambiguity in the definition
of $\hat C_3$ in that  $\hat C_3\to  \hat C_3 + \hat A_1\wedge \hat B_2$
changes the form of the action and the form of the gauge transformations. It
is this second formulation that we use in section \ref{NSflux} where we turn
on NS three-form flux.}
In these conventions (\ref{S10'})
is invariant under the following 
three Abelian gauge transformations
\begin{eqnarray}
  \label{gtrIIA}
  \delta \hat A_1 &=&  d  \hat\Theta \ ,\qquad 
  \delta \hat C_3 = d \hat \Sigma_2\ , \\
  \delta \hat B_2 &=& d  \hat\Lambda_1 \ , 
\quad  \delta\hat C_3 = \hat A_1 \wg d \hat \Lambda_1   \ .\nn
\end{eqnarray}

In order to compactify the action (\ref{S10'}) on a \CY threefold $Y_3$
we expand the ten-dimensional fields in harmonic forms on ${Y_3}$. 
As already reviewed in section~\ref{RRf}  the K\"ahler
class deformations of the metric are in one to one correspondence with the harmonic 
$(1,1)$-forms while the complex structure deformations are in one to one
correspondence with the harmonic $(1,2)$-forms on ${Y_3}$. 
Furthermore, the forms $\hat A_1, \hat B_2, \hat C_3$ are expanded 
according to 
\begin{eqnarray}
  \label{CYred}
  \hat A_1 & = & A^0 \ ,\nn \\
  \hat B_2 & = & B_2 + b^i\, \ox_i \ ,\\
  \hat C_3 & = & C_3 + A^i \wedge \ox_i + \xi^A \ax_A + \tilde\xi_B \beta^B\ , \nn
\end{eqnarray}
where as reviewed in appendix \ref{CYM},
$\ox_i \, , i = 1, \ldots, h_{1,1}$ are harmonic $(1,1)$ forms on ${Y_3}$
and $(\ax_A,\bx^A), A=0, \ldots, h_{1,2} $ is a real basis on $H^3({Y_3})$.
$A^0$ is the graviphoton and together
with the graviton $g_{\mu \nu}$ describe the bosonic components of
the gravitational multiplet.
The other $h_{1,1}$ vector fields $A^i$
combine with the $t^j = b^j +i v^j$ into
$h_{1,1}$ vector-multiplets. 
The $h_{1,2}$ complex structure deformations $z^a$ together with 
$\xi^a, \tilde\xi_a$ form $h_{1,2}$ hypermultiplets and 
$B_2, \phi, \xi^0, \tilde\xi_0$ form a tensor multiplet.
In $d=4$ the two-form $B_2$ can be dualized to a scalar and hence
the tensor multiplet can be turned into an additional (universal)
hypermultiplet.
$C_3$ carries no degrees of
freedom in $d=4$  but is dual to a constant.

Using (\ref{K}) (\ref{A-N}) and  (\ref{CYred}) the various terms in the
Lagrangian integrated over the \CY space become 
\begin{eqnarray}
  \label{HH}
  - \frac14 \int_{Y_3}\hat H_3 \awedge \hat H_3 
  & = & - \frac{\K}{4} \, d B_2 \awedge d B_2 - \K g_{ij} db^i ´\awedge db^j \
  ,\nn \\ 
  \nn \\
  - \frac12 \int_{Y_3} \hat F_2 \awedge \hat F_2 & = & - \frac{\K}{2} \, d A^0
  \awedge dA^0 \ ,\\ 
  \nn \\
  - \frac12 \int_{Y_3} {\hat {F}_4} \awedge {\hat {F}_4} & = & - \frac{\K}{2}
  \, (d C_3 - d A^0\wedge B_2) \awedge (d C_3 - d A^0\wg B_2) \nn \\
  & & - 2 \K g_{ij} (d A^i - d A^0 b^i) \awedge (d A^j - d A^0 b^j) \nn \\
  & & + \frac{1}{2}\left(\IM \Mc ^{-1} \right)^{AB} 
  \Big[ d\tilde\xi_A +  \Mc_{AC} d\xi^C \Big] \awg \Big[ d\tilde\xi_B + 
  \bar \Mc_{BD} d\xi^D \Big] \, , \nn
\end{eqnarray}
where $g_{ij}$  and $\K$ were defined in (\ref{K}),(\ref{gH11}). 
Finally, for the topological terms we find 
\begin{eqnarray}
\label{CS}
\int_{Y_3}
 \mathcal {L}_{top} & = & - \frac12 \Big[B_2 \wedge 
d(\tilde\xi_A d \xi^A -\xi^A d\tilde\xi_A) + b^i dA^j \wedge
  dA^k \K_{ijk} \nn \\
  & & \qquad \quad - b^i b^j d A^k \wg dA^0 \K_{ijk} + \frac13 b^i b^j  b^k d
 A^0 \wg d A^0 \K_{ijk}\Big] \, .
\end{eqnarray}
Defining the four-dimensional dilaton $\phi$ via 
$e^{-2\phi} = e^{-2\hat\phi} \K$
the action in $d=4$ becomes\footnote{%
Strictly speaking also the K\"ahler moduli $t^i$ have to be redefined
by a dilaton dependent factor \cite{BCF}. In order not to overload the
notation we use the same $t^i$ also for the redefined moduli.} 
\begin{eqnarray}
  \label{S4}
  S & = & \int e^{-2\phi} \left( \frac12 R ^\ast \! {\bf 1} + 2d \phi
  \awedge d \phi - \frac14 H_3 \awedge H_3 -
  g_{ij} dt^i \awg d {\bar t}^j - g_{ab} d z^a \awg d {\bar z}^b \right) \nn \\ 
  & & - \frac12 \int \Big[ \K F_2 \awedge F_2 + 4 \K g_{ij} (d A^i - d A^0 b^i)
  \awedge (d A^j - d A^0 b^j) \Big] \nn \\ 
  & & + \frac{1}{2}\left(\IM \Mc ^{-1} \right)^{AB} 
  \Big[ d\tilde\xi_A +  \Mc_{AC} d\xi^C \Big] \awg \Big[ d\tilde\xi_B + 
  \bar \Mc_{BD} d\xi^D \Big] \nn\\
  & & +\frac12 \int H_3 \wedge (\tilde\xi_A d \xi^A -\xi^A d\tilde\xi_A)  \\
  & & - \frac12 \int \left[ b^i dA^j \wedge dA^k - b^i b^j d A^k \wg d A^0 +
  \frac13 b^i b^j b^k d A^0 \wg d A^0 \right] \K_{ijk} \nn \\
  & & -\frac12 \int \K(dC_3 - d A^0\wedge B_2) \awedge (dC_3 - d A^0\wedge
  B_2)\ . \nn
\end{eqnarray}

The next step is the dualization of $C_3$ following appendix~\ref{3f}.
Using the results in this appendix we find that the dual of the Lagrangian
\begin{equation}
  \label{A3}
  \L_{C_3} = - \frac{\K}{2} (dC_3 - d A^0 \wg B_2) \awg (dC_3 - d A^0 \wg B_2) 
\end{equation}
is given by 
\begin{equation}
  \label{A3dual}
  \L_{e_0} =  - \frac{1}{2 \, \K} e_0^2 \ast {\bf 1} - 
  e_0 dA^0 \wg B_2 \, ,
\end{equation}
with $e_0$ being an arbitrary constant parameter.
Replacing (\ref{A3}) by (\ref{A3dual}) in the action (\ref{S4})
and collecting the terms involving $H_3$ we obtain (after partial integration)
\begin{equation}
  \label{actH}
  \L_{H_3} =  \left[ - \frac{1}{4} e^{-2\phi}H_3 \awg H_3  + \frac12 H_3 \wg 
  \left(\tilde\xi_A d \xi^A -\xi^A d \tilde\xi_A \right) + 
  e_0 H_3 \wg A^0 \right]
  \, . 
\end{equation}
We see that $e_0$ induces a  
four-dimensional Green-Schwarz term $H_3 \wg A^0$ into action.
The standard form of the type IIA action is obtained by also dualizing $B_2$
to an axionic scalar $a$.
For details we refer the reader to appendix \ref{Bdual} while here we only
record the final result
\begin{equation}
  \label{dact}
  \L_{H_3}\to \L_a = - \frac{e^{2\phi}}{4} \, 
  \Big[ Da + 
  (\tilde\xi_A d \xi^A -\xi^A d\tilde\xi_A)
  \Big] \wg \rule{0pt}{12pt}^* \! 
  \Big[ Da + 
  (\tilde\xi_A d\xi^A-\xi^A d\tilde\xi_A) \Big] \ ,
\end{equation}
where
\begin{equation}
  \label{cda}
  D a = da + 2 e_0 A^0 \ .
\end{equation}
As anticipated the axionic scalar $a$ is charged under a local Peccei-Quinn
gauge symmetry with $e_0$ being the gauge charge. Thus even
ordinary Calabi-Yau compactifications of type IIA give a one-parameter family
of four-dimensional effective theories which are generically gauged
rather then ordinary supergravities. For  $e_0 = 0$ 
one recovers the standard type IIA
supergravity of ref.\ \cite{BCF}.

The final task of this appendix is to rewrite the action in the form of
standard gauged supergravity \cite{N=2}. To do this we use the dual
action (\ref{dact}) instead of (\ref{actH}) and perform 
the Weyl rescaling $g_{\mu \nu} \to e^{2\phi} g_{\mu \nu}$ 
in order to go to the Einstein frame.
Furthermore the scalars $(\phi,a, z^a, \xi^A,\tilde\xi_A)$ which together 
form $h_{1,2} +1$ hypermultiplets are denoted collectively
by $q^u$.
In these variables the action (\ref{S4}) reads 
\begin{equation}
    \label{S4f}
  S =  \int \frac12 R ^* {\bf 1} - g_{ij} dt^i \awg d
  {\bar t}^j - h_{uv} Dq^u \awg Dq ^v + \frac{1}{2}\, \IM \N_{IJ} F^I\awg F^{J}
  + \frac{1}{2} \, \RE \N_{IJ} F^I \wg F^J - V_E  \ .
\end{equation}
where 
\begin{eqnarray}
  \label{qkt}
  h_{uv} Dq^u \awg Dq ^v& = &  d\phi \awg d\phi + g_{ab} dz^a \awg dz^b
  \\  
  & & + \frac{e^{4\phi}}{4} \, \Big[ Da + 
  (\tilde\xi_A d \xi^A-\xi^A
  d\tilde\xi_A) \Big] \wg \rule{0pt}{12pt}^* \Big[ Da + 
  (\tilde\xi_A d \xi^A-\xi^A d\tilde\xi_A) \Big] \nn \\
  & & - \frac{e^{2\phi}}{2}\left(\IM \Mc^{-1} \right)^{AB} 
  \Big[ d\tilde\xi_A + \Mc_{AC} d\xi^C \Big]
   \wg \rule{0pt}{12pt}^* \Big[ d\tilde\xi_B + \bar \Mc_{BD} d\xi^D \Big]  \, ,\nn
\end{eqnarray}
and $\Mc_{AB}$ was defined in (\ref{N_G}).
In  ref.\ \cite{FeS}
it was shown in that $ h_{uv}$ is a quaternionic metric 
in accord with the constraints of $N=2$ supergravity
that the scalars in the hypermultiplets span a quaternionic manifold.

Using (\ref{gH11}) it is straightforward to show that the  gauge couplings  
in (\ref{S4f}) are given by (\ref{eq:N}).
Finally the potential $V_E$ 
in (\ref{S4f}) is  given by  
\begin{equation}
V_E = 
\frac{e^{4\phi}}{2 \K}\ e_0^{2 \; *} {\bf 1}\ .
\end{equation}
This potential coincides with (\ref{ApotE}) for $e_i=0$ and
thus the consistency with gauged supergravity
can be shown analogously to section 2.3.

\section{Poincar\'e dualities}
\setcounter{equation}{0}

For an arbitrary $p$-form in $d$ dimensions one always has the choice to
describe the action in terms of a Poincar\'e dual form. The nature of
the dual form differs in the massless and massive case.\footnote{We thank
F.\ Quevedo for educating us on this subject.}
A massless $p$-form in $d$ dimensions describes 
$\binom{d-2}{p}$ physical degrees of freedom
 while a massive  $p$-form in $d$ dimensions contains 
$\binom{d-1}{p}$  
degrees of freedom. The difference can be easily understood from
a generalized Higgs mechanism where a $p$-form 
`eats' a massless $p-1$-form and thus the number of degrees of freedom
change by 
$\binom{d-2}{p-1}$.
Therefore a massless $p$-form in $d$ dimensions
is dual to a $(d-p-2)$-form while a massive $p$-form
is dual to a $(d-p-1)$-form.
A massless ($d-1$)-form is special in that it is dual to a constant.
In $d=4$ this implies that a massless three-form is dual to a constant,
a massless two-form is dual to a scalar while a massive 2-from
is dual to a vector (a 1-form). Let us discuss these cases in turn.

\subsection{Dualization of a massless $B_2$}
\label{Bdual}
Let us first consider the dualization of a massless two-form $B_2$
with field strength $H_3 = d B_2$ to a scalar $a$. We start from the generic
action 
\begin{equation}
  \label{actHg}
  S_{H_3} = -\int \Big[\frac{g}{4} H_3 \awg H_3  - \frac12 H_3 \wg J_1\Big]
  \, ,
\end{equation}
where $g$ is an arbitrary function of the scalars while 
$J_1$ is a generic 1-form depending on the scalars and possibly
some gauge field $A_1$. The
dualization can be carried out  by introducing a scalar field $a$ as a Lagrange
multiplier and adding the term $H_3 \wg d a$ to $S_{H_3}$.
Treating $H_3$ as an independent three-form (not being $dB_2$)
the equation of motion for $a$ implies  $H_3=dB_2$ while the equation
of motion for $H_3$ reads $ \rule{0pt}{8pt}^* H_3 = \frac{1}{g} (da + J_1)$.
Inserted back into the action (\ref{actHg}) one obtains  the dual action
\begin{equation}
  \label{Bdact}
  S_a = - \int \frac1{4g} (da + J_1) \awg (da + J_1)
  \ .
\end{equation}

There is an another way of treating the dualizations which turns
out to be useful in understanding the dualization of a three-form in four
dimensions. Consider the equation of motion for $B_2$ 
\begin{equation}
  d (g \, \rule{0pt}{8pt}^* H_3 - J_1) = 0,
\end{equation}
which can be derived from (\ref{actHg}).
It is solved by
\begin{equation}
  g \rule{0pt}{8pt}^* H_3 - J_1 = d a \, ,
\end{equation}
with $a$ being some arbitrary scalar field. 
The equation of motion for this field
is dictated by the Bianchi identity of $H_3$
\begin{equation}
  0 = d H_3 = d \left [\frac{1}{g} \, \rule{0pt}{8pt}^*(d a + J_1) \right] \, , 
\end{equation}
which in turn can be obtained from the action (\ref{Bdact}).
This implies
that the two ways described for the dualization of $B_2$ are equivalent.

\subsection{Dualization of the three-form}
\label{3f}

Next we consider the dualization of a three-form in 4 dimensions. We start from a
generic action for a three-form $C_3$ possibly coupled 
to two-forms, 1-forms and scalars 
\begin{equation}
  \label{actA3}
  S_{C_3} = - \int \Big[\frac{g}{4}(d C_3 - J_4)\awg (d C_3-J_4) + 
  \frac{h}{2} d C_3 \Big] \, , 
\end{equation}
where $g,h$  denote two arbitrary scalar functions and $J_4$ is
an 4-form which can depend on the two-forms, 1-forms and scalars
present in the spectrum. 

For the field strength of a three-form in 4 dimensions there is no proper Bianchi
identity since no 5-forms exist. That is why the second way of dualizing forms
presented in the previous section, by exchanging the equation of motion with
the Bianchi identity, can not work in this case. The only consistent way to
proceed is to add a Lagrange multiplier to the action (\ref{actA3}) \cite{BGG}
\begin{equation}
  \label{actA3'}
  S_{C_3} = - \int \Big[\frac{g}{4}(d C_3 - J_4)\awg (d C_3-J_4) + 
  \frac{h}{2} d C_3 + \frac{e_0}{2} d C_3 \Big] \, , 
\end{equation}
where $e_0$  is a constant.
The equation of motion for $dC_3$ imply
\begin{equation}
  \frac{g}{2} \; \rule{0pt}{10pt}^* (dC_3 - J_4)  = -
  \frac{h + e_0}{2} \, . 
\end{equation}
Inserted back into the action (\ref{actA3}) and using 
$^{**} dC_3 = -dC_3$ one obtains
\begin{equation}
  S_{e_0} = - \int \left[\frac{1}{4g}(h + e_0)^{2 \;*} {\bf 1} +
  \frac12 (h + e_0) J_4 \right]\ .
\end{equation}
As we see a potential for the scalar fields is induced and $e_0$
play the role of a cosmological constant.

\subsection{Dualization of a massive two-form}
\label{Mdual}

Finally, let us dicuss the dualization of the a massive two-form $B_2$ 
\cite{Tap,QuT,SmS}.
We start from a generic action
\begin{equation}
  \label{SB}
  S_{B_2} = - \int \left[ g H_3 \awg H_3 + M^2 B_2 \awg B_2 + M_T^2 B_2  \wg
  B_2 + B_2 \wg J_2 \right] \, ,
\end{equation}
where $g,M,M_T$ can be field dependent couplings and $J_2$
is a two-form which can depend on the gauge potential $A_1$ and/or
some scalar fields. ($J_2$  does not depend on $B_2$.)
We can  treat $B_2$ and
$H_3$ as independent fields and ensure $H_3 = d B_2$ by the equations
of motion. This is achieved in the action
\begin{equation}
  \label{SB'}
  S'_{B_2} = - \int \left[ - g H_3 \awg H_3 + 2 g H_3 \awg d B_2 + M^2 B_2 \awg
  B_2 + M_T^2 B_2  \wg B_2 + B_2 \wg J_2 \right] \ ,
\end{equation}
which indeed has $H_3 = d B_2$ as the equation of motion for $H_3$.
So by inserting $H_3 = d B_2$ into (\ref{SB'}) we obtain (\ref{SB}). 
On the other
hand one can eliminate $B_2$ through its equation of motion and obtain an
action expressed only in terms of $H_3$. The equation of motion for $B_2$ from
(\ref{SB'}) is
\begin{equation}
  \label{eomB}
  2 M^{2 \; *} B_2 + 2M_T^2 B_2 + J_2 - 2 d^{\;*}(gH_3) = 0\ ,
\end{equation}
which is solved by
\begin{eqnarray}
  \label{B(H)}
  ^*B_2 & = & \frac{1}{M^4 + M_T^4} \Big[ M^2 d^{\;*}(g H_3) + M_T^{2 \; *}
  d^{\;*}(g H_3) - \frac{M^2}{2} J_2 - \frac{M_T^2}{2} \, \rule{0pt}{10pt}^*
  \! J_2 \Big]\, \quad  \textrm{or} \nn \\
  B_2 & = & \frac{1}{M^4 + M_T^4} \Big[ M_T^2 d^{\;*} (g H_3) - M^{2\; *}
  d^{\;*}(g H_3)+ \frac{M^2}{2} \, \rule{0pt}{10pt}^* \! J_2 - \frac{M_T^2}{2}
  J_2 \Big]\  .
\end{eqnarray}
Inserted back into the action (\ref{SB'}) results in 
\begin{eqnarray}
  \label{SB"}
  S^{''}_{B_2} & = &  \int \bigg[ g H_3 \awg H_3 - \frac{M^2}{M^4 + M_T^4}
  \left(d^{\,*}(g H_3) - \frac12 J_2 \right) \awg \left(d^{\,*}(g H_3) -
  \frac12 J_2 \right) \nn \\ 
  & & \qquad + \frac{M_T^2}{M^4 + M_T^4} \left(d^{\,*}(g H_3) -
  \frac12 J_2 \right) \, \wg \, \left(d^{\,*}(g H_3) - \frac12 J_2 \right)
  \bigg] \, .
\end{eqnarray}
We can now replace $H_3$ by its Poincar\'e dual 
one-form $A^H =g  \, ^*H_3$ and the dual
action for the massive field $A^H$ is
\begin{eqnarray}
  \label{SA}
  S_{A^H} & = & - \int \Bigg[\frac{1}{g} A^H \awg A^H + \frac{M^2}{M^4 + M_T^4}
  \left(d A^H - \frac12 J_2 \right) \awg \left(d A^H - \frac12 J_2 \right) \nn \\ 
  & & \nn \\
  & & \hspace{2.5cm}  - \frac{M_T^2}{M^4 + M_T^4} \left(d A^H - \frac12 J_2
  \right) \, \wg \, \left(d A^H - \frac12 J_2 \right) \Bigg] \, .
\end{eqnarray}
As promised this is the action for a massive one-form $A^H$.
\vskip 1cm

{\large \bf Acknowledgments}

This work is supported by DFG -- The German Science Foundation,
GIF -- the German--Israeli Foundation for Scientific Research,
the European RTN Program HPRN-CT-2000-00148 and the
DAAD -- the German Academic Exchange Service.

We have greatly benefited from conversations with 
L.~Andrianopoli, J.-P.~Derendinger, 
B.~de Wit, R.~Grimm, B.~Gunara, S.~Gurrieri,  D.~L\"ust, 
P.~Mayr, T.~Mohaupt, F.~Quevedo,
H.~Singh, A.~Strominger, C.~Vafa, S.~Vandoren, 
A.~Van Proeyen,  and  M.~Zagermann.


\providecommand{\href}[2]{#2}
\end{document}